\documentclass[twocolumn,twocolappendix]{aastex631}%linenumbers
\graphicspath{{fig/}}
\usepackage{paracol}
\usepackage{threeparttablex}
\usepackage{color}

\usepackage{multirow} % 用于跨多行的单元格
\usepackage{textcomp} % 在导言区加入此行以使用 textcomp 包
\usepackage{natbib}

\usepackage{amsmath}

\begin{document}

\title{EP240801a/XRF 240801B: An X-ray Flash Detected by the Einstein Probe \\ and Implications of its Multiband Afterglow}

\correspondingauthor{Dong Xu, Wei-Hua Lei}
\email{dxu@nao.cas.cn, leiwh@hust.edu.cn}

\author[0009-0001-8155-7905]{Shuai-Qing Jiang}
\affiliation{National Astronomical Observatories, Chinese Academy of Sciences, Beijing 100101, People's Republic of China}
\affiliation{School of Astronomy and Space Science, University of Chinese Academy of Sciences, Chinese Academy of Sciences, Beijing 100049, People's Republic of China}

\author[0000-0003-3257-9435]{Dong Xu}
\affiliation{National Astronomical Observatories, Chinese Academy of Sciences, Beijing 100101, People's Republic of China}
\affiliation{Altay Astronomical Observatory, Altay, Xinjiang 836500, People's Republic of China}

\author[0009-0005-5404-2745]{Agnes P. C. van Hoof}
\affiliation{Radboud University Nijmegen, Department of Astrophysics/IMAPP, P.O.~Box 9010, Nijmegen, 6500~GL, The Netherlands}

\author[0000-0003-3440-1526]{Wei-Hua Lei}
\affiliation{Department of Astronomy, School of Physics, Huazhong University of Science and Technology, Wuhan, 430074, People’s Republic of China}

\author{Yuan Liu}
\affiliation{National Astronomical Observatories, Chinese Academy of Sciences, Beijing 100101, People's Republic of China}

\author[0000-0003-2915-7434]{Hao Zhou}
\affiliation{Purple Mountain Observatory, Chinese Academy of Sciences, Nanjing 210023, People's Republic of China}

\author[0000-0001-9834-2196]{Yong Chen}
\affiliation{Key Laboratory of Particle Astrophysics, Institute of High Energy Physics, Chinese Academy of Sciences, Beijing 100049, People’s Republic of China}

\author[0009-0002-7730-3985]{Shao-Yu Fu}
\affiliation{National Astronomical Observatories, Chinese Academy of Sciences, Beijing 100101, People's Republic of China}

\author[0000-0002-5485-5042]{Jun Yang}
\affiliation{School of Astronomy and Space Science, Nanjing University, Nanjing 210093, People's Republic of China}
\affiliation{Key Laboratory of Modern Astronomy and Astrophysics (Nanjing University), Ministry of Education, People's Republic of China}

\author{Xing Liu}
\affiliation{National Astronomical Observatories, Chinese Academy of Sciences, Beijing 100101, People's Republic of China}
\affiliation{School of Astronomy and Space Science, University of Chinese Academy of Sciences, Chinese Academy of Sciences, Beijing 100049, People's Republic of China}

\author[0000-0002-9022-1928]{Zi-Pei Zhu}
\affiliation{National Astronomical Observatories, Chinese Academy of Sciences, Beijing 100101, People's Republic of China}

\author[0000-0003-3460-0103]{Alexei~V.~Filippenko}
\affiliation{Department of Astronomy, University of California, Berkeley, CA 94720-3411, USA}

\author[0000-0001-5679-0695]{Peter G. Jonker}
\affiliation{Radboud University Nijmegen, Department of Astrophysics/IMAPP, P.O.~Box 9010, Nijmegen, 6500~GL, The Netherlands}

\author[0000-0001-9435-1327]{A. S. Pozanenko}
\affiliation{Space Research Institute, Russian Academy of Sciences, Profsoyuznaya ul. 84/32, Moscow, 117997, Russia}
\affiliation{National Research University “Higher School of Economics”, Myasnitskaya ul. 20, Moscow, 101000, Russia}
\affiliation{Moscow Institute of Physics and Technology (MIPT), Institutskiy Pereulok, 9, Dolgoprudny, 141701, Russia}

\author[0000-0002-3100-6558]{He Gao}
\affiliation{School of Physics and Astronomy, Beijing Normal University, Beijing 100875, People's Republic of China}
\affiliation{Institute for Frontier in Astronomy and Astrophysics, Beijing Normal University, Beijing 102206, People's Republic of China}

\author[0000-0002-6299-1263]{Xue-Feng Wu}
\affiliation{Purple Mountain Observatory, Chinese Academy of Sciences, Nanjing 210023, People's Republic of China}

\author[0000-0002-9725-2524]{Bing Zhang}
\affiliation{Nevada Center for Astrophysics, University of Nevada Las Vegas, NV 89154, USA}
\affiliation{Department of Physics and Astronomy, University of Nevada Las Vegas, NV 89154, USA}

\author[0000-0001-5169-4143]{Gavin P Lamb}
\affiliation{Astrophysics Research Institute, Liverpool John Moores University, IC2 Liverpool Science Park, 146 Brownlow Hill, Liverpool L3 5RF, United Kingdom}

\author[0000-0002-4036-7419]{Massimiliano De Pasquale}
\affiliation{University of Messina, MIFT Department, Via F.S. D'Alcontres 31, Messina, 98166, Italy}

\author[0000-0001-7946-4200]{Shiho Kobayashi}
\affiliation{Astrophysics Research Institute, Liverpool John Moores University, IC2 Liverpool Science Park, 146 Brownlow Hill, Liverpool L3 5RF, United Kingdom}

\author[0000-0002-8686-8737]{Franz Erik Bauer}
\affiliation{Instituto de Alta Investigaci{\'{o}}n, Universidad de Tarapac{\'{a}}, Casilla 7D, Arica, Chile}

\author[0000-0002-9615-1481]{Hui Sun}
\affiliation{National Astronomical Observatories, Chinese Academy of Sciences, Beijing 100101, People's Republic of China}

\author[0000-0003-3457-9375]{Giovanna Pugliese}
\affiliation{Anton Pannekoek Institute of Astronomy, University of Amsterdam, Science Park 904, 1098 XH Amsterdam, The Netherlands}

\author{Jie An}
\affiliation{National Astronomical Observatories, Chinese Academy of Sciences, Beijing 100101, People's Republic of China}
\affiliation{School of Astronomy and Space Science, University of Chinese Academy of Sciences, Chinese Academy of Sciences, Beijing 100049, People's Republic of China}

\author[0000-0003-3703-4418]{Valerio D'Elia}
\affiliation{Space Science Data Center (SSDC) - Agenzia Spaziale Italiana (ASI), I-00133 Roma, Italy}
\affiliation{Osservatorio Astronomico di Roma, via Frascati 33, I-00040 Monte Porzio Catone, Italy}

\author[0000-0002-8149-8298]{Johan P. U. Fynbo}
\affiliation{The Cosmic Dawn Centre (DAWN), Denmark}
\affiliation{Niels Bohr Institute, University of Copenhagen, Jagtvej 155, DK-2200, Copenhagen N, Denmark}

\author[0000-0002-2636-6508]{WeiKang Zheng}
\affiliation{Department of Astronomy, University of California, Berkeley, CA 94720-3411, USA}

\author{Alberto J. C. Tirado}
\affiliation{Instituto de Astrof\'isica de Andaluc\'ia (IAA-CSIC), Glorieta de la Astronom\'ia s/n, 18008 Granada, Spain}
\affiliation{Ingeniería de Sistemas y Autom\'atica, Universidad de M\'alaga, Unidad Asociada al CSIC por el IAA, Escuela de Ingenier\'ias Industriales, Arquitecto Francisco Pe\~nalosa, 6, Campanillas, 29071 M\'alaga, Spain}

\author[0000-0002-5596-5059]{Yi-Han Iris Yin}
\affiliation{School of Astronomy and Space Science, Nanjing University, Nanjing 210093, People's Republic of China}
\affiliation{Key Laboratory of Modern Astronomy and Astrophysics (Nanjing University), Ministry of Education, People's Republic of China}

\author[0000-0002-5400-3261]{Yuan-Chuan Zou}
\affiliation{Department of Astronomy, School of Physics, Huazhong University of Science and Technology, Wuhan, 430074, People’s Republic of China}

\author[0000-0001-9434-3837]{Adam~T.~Deller}
\affiliation{Center for Astrophysics and Supercomputing, Swinburne University of Technology, P.O. Box 218, Hawthorn, Vic 3122, Australia}
\affiliation{ARC Centre of Excellence for Gravitational Wave Discovery (OzGrav), Hawthorn, Victoria, 3122, Australia}

\author{N. S. Pankov}
\affiliation{Space Research Institute, Russian Academy of Sciences, Profsoyuznaya ul. 84/32, Moscow, 117997, Russia}
\affiliation{National Research University “Higher School of Economics”, Myasnitskaya ul. 20, Moscow, 101000, Russia}

\author{A. A. Volnova}
\affiliation{Space Research Institute, Russian Academy of Sciences, Profsoyuznaya ul. 84/32, Moscow, 117997, Russia}

\author{A. S. Moskvitin}
\affiliation{Space Research Institute, Russian Academy of Sciences, Profsoyuznaya ul. 84/32, Moscow, 117997, Russia}

\author{O. I. Spiridonova}
\affiliation{Space Research Institute, Russian Academy of Sciences, Profsoyuznaya ul. 84/32, Moscow, 117997, Russia}

\author{D. V. Oparin}
\affiliation{Space Research Institute, Russian Academy of Sciences, Profsoyuznaya ul. 84/32, Moscow, 117997, Russia}

\author{V. Rumyantsev}
\affiliation{Crimean astrophysical observatory, 298409, Nauchny, Crimea}

\author{O. A. Burkhonov}
\affiliation{Ulugh Beg Astronomical Institute, Uzbek Academy of Sciences, Tashkent, Uzbekistan}

\author{Sh.A. Egamberdiyev}
\affiliation{Ulugh Beg Astronomical Institute, Uzbek Academy of Sciences, Tashkent, Uzbekistan}
\affiliation{National University of Uzbekistan, Tashkent, Uzbekistan}

\author{V. Kim}
\affiliation{Fesenkov Astrophysical Institute, Almaty, 050020, Kazakhstan}

\author{M. Krugov}
\affiliation{Fesenkov Astrophysical Institute, Almaty, 050020, Kazakhstan}

\author{A. M. Tatarnikov}
\affiliation{Sternberg Astronomical Institute, Lomonosov Moscow State University, Moscow, 119191 Russia}

\author{R. Inasaridze}
\affiliation{Evgeni Kharadze Georgian National Astrophysical Observatory, 0301, Adigeni, Abastumani, Mt. Kanobili, Georgia}
\affiliation{Samtskhe-Javakheti State University, Rustaveli Str. 113, Akhaltsikhe, 0080, Georgia}

\author[0000-0001-7821-9369]{Andrew J. Levan}
\affiliation{Radboud University Nijmegen, Department of Astrophysics/IMAPP, P.O.~Box 9010, Nijmegen, 6500~GL, The Netherlands}
\affiliation{University of Warwick, Department of Physics, Coventry, CV4 7AL, United Kingdom}

\author[0000-0002-7517-326X]{Daniele Bj\o{}rn Malesani}
\affiliation{The Cosmic Dawn Centre (DAWN), Denmark}
\affiliation{Niels Bohr Institute, University of Copenhagen, Jagtvej 155, DK-2200, Copenhagen N, Denmark}

\author[0000-0003-3193-4714]{Maria E. Ravasio}
\affiliation{Radboud University Nijmegen, Department of Astrophysics/IMAPP, P.O.~Box 9010, Nijmegen, 6500~GL, The Netherlands}
\affiliation{Osservatorio Astronomico di Brera, Istituto Nazionale di Astrofisica, Merate 23807, Italy}

\author[0000-0001-8602-4641]{Jonathan Quirola-V\'asquez}
\affiliation{Radboud University Nijmegen, Department of Astrophysics/IMAPP, P.O.~Box 9010, Nijmegen, 6500~GL, The Netherlands}

\author[0009-0007-6927-7496]{Joyce N. D. van Dalen}
\affiliation{Radboud University Nijmegen, Department of Astrophysics/IMAPP, P.O.~Box 9010, Nijmegen, 6500~GL, The Netherlands}

\author[0000-0003-2276-4231]{Javi S\'anchez-Sierras}
\affiliation{Radboud University Nijmegen, Department of Astrophysics/IMAPP, P.O.~Box 9010, Nijmegen, 6500~GL, The Netherlands}

\author[0000-0003-0245-9424]{Daniel Mata S\'anchez}
\affiliation{Instituto de Astrof\'isica de Canarias, E-38205, La Laguna, Tenerife, Spain}
\affiliation{Universidad de La Laguna, Departamento de Astro\'isica,  E-38206, La Laguna, Tenerife, Spain}

\author[0000-0001-7221-855X]{Stuart P. Littlefair}
\affiliation{University of Sheffield, Department of Physics and Astronomy, Sheffield, S3 7RH, United Kingdom}

\author[0009-0000-6374-3221]{Jennifer A. Chac\'on}
\affiliation{Instituto de Astrof{\'{\i}}sica, Facultad de F{\'{i}}sica, Pontificia Universidad Cat{\'{o}}lica de Chile, Campus San Joaquín, Av. Vicuña Mackenna 4860, Macul Santiago, Chile, 7820436}
\affiliation{Millennium Institute of Astrophysics, Nuncio Monse{\~{n}}or S{\'{o}}tero Sanz 100, Of 104, Providencia, Santiago, Chile}

\author[0000-0002-5297-2683]{Manuel A. P. Torres}
\affiliation{Instituto de Astrof\'isica de Canarias, E-38205, La Laguna, Tenerife, Spain}
\affiliation{Universidad de La Laguna, Departamento de Astro\'isica,  E-38206, La Laguna, Tenerife, Spain}

\author[0000-0001-9842-6808]{Ashley A. Chrimes}
\affiliation{European Space Agency (ESA), European Space Research and Technology Centre (ESTEC), Keplerlaan 1, Noordwijk, 2201 AZ, The Netherlands}
\affiliation{Radboud University Nijmegen, Department of Astrophysics/IMAPP, P.O.~Box 9010, Nijmegen, 6500~GL, The Netherlands}

\author[0000-0003-2700-1030]{Nikhil Sarin}
\affiliation{The Oskar Klein Centre, Department of Physics, Stockholm University, AlbaNova, Stockholm, SE-106 91, Stockholm, Sweden}
\affiliation{Nordita, Stockholm University and KTH Royal Institute of Technology, Hannes Alfv\'ens v\"ag 12, Stockholm, SE-106 91, Stockholm, Sweden}

\author[0000-0001-5108-0627]{Antonio Martin-Carrillo}
\affiliation{School of Physics and Centre for Space Research, University College Dublin, Dublin, D04
V1W8, Dublin, Ireland}

\author[0000-0003-4236-9642]{Vik Dhillon}
\affiliation{University of Sheffield, Department of Physics and Astronomy, Sheffield, S3 7RH, United Kingdom}

\author[0000-0002-6535-8500]{Yi Yang}
\affiliation{Physics Department, Tsinghua University, Beijing, 100084, People’s Republic of China}

\author[0000-0001-5955-2502]{Thomas~G.~Brink}
\affiliation{Department of Astronomy, University of California, Berkeley, CA 94720-3411, USA}

\author[0000-0002-3324-4824]{Rebecca~L.~Davies}
\affiliation{Center for Astrophysics and Supercomputing, Swinburne University of Technology, P.O. Box 218, Hawthorn, Vic 3122, Australia}

\author[0000-0002-2898-6532]{Sheng Yang}
\affiliation{Institute for Gravitational Wave Astronomy, Henan Academy of Sciences, Zhengzhou 450046, Henan, People’s Republic of China}

\author[0000-0002-9928-0369]{Amar Aryan}
\affiliation{Graduate Institute of Astronomy, National Central University, 300 Jhongda Road, 32001 Jhongli, Taiwan}

\author[0000-0002-1066-6098]{Ting-Wan Chen}
\affiliation{Graduate Institute of Astronomy, National Central University, 300 Jhongda Road, 32001 Jhongli, Taiwan}

\author[0000-0002-5105-344X]{Albert K. H. Kong}
\affiliation{Institute of Astronomy, National Tsing Hua University, No. 101 Sect. 2 Kuang-Fu Road, 30013 Hsinchu, Taiwan}

\author[0000-0002-0096-3523]{Wen-Xiong Li}
\affiliation{National Astronomical Observatories, Chinese Academy of Sciences, Beijing 100101, People's Republic of China}

\author[0000-0002-4205-0933]{Rui-Zhi Li}
\affiliation{Yunnan Observatories, Chinese Academy of Sciences, Kunming 650216, People's Republic of China}
\affiliation{School of Astronomy and Space Science, University of Chinese Academy of Sciences, Chinese Academy of Sciences, Beijing 100049, People's Republic of China}

\author[0000-0002-7077-7195]{Jirong Mao}
\affiliation{Yunnan Observatories, Chinese Academy of Sciences, Kunming 650216, People's Republic of China}
\affiliation{Center for Astronomical Mega-Science, Chinese Academy of Sciences, Beijing 100012, People's Republic of China}
\affiliation{Key Laboratory for the Structure and Evolution of Celestial Objects, Chinese Academy of Sciences, Kunming 650216, People's Republic of China}

\author{Ignacio P\'erez-Garc\'ia}
\affiliation{Instituto de Astrof\'isica de Andaluc\'ia (IAA-CSIC), Glorieta de la Astronom\'ia s/n, 18008 Granada, Spain}

\author{Emilio J. Fern\'andez-Garc\'ia}
\affiliation{Instituto de Astrof\'isica de Andaluc\'ia (IAA-CSIC), Glorieta de la Astronom\'ia s/n, 18008 Granada, Spain}

\author{Moira Andrews}
\affiliation{Las Cumbres Observatory, 6740 Cortona Drive, Suite 102, Goleta, CA 93117-5575, USA}

\author[0000-0003-4914-5625]{Joseph Farah}
\affiliation{Las Cumbres Observatory, 6740 Cortona Drive, Suite 102, Goleta, CA 93117-5575, USA}
\affiliation{Department of Physics, University of California, Santa Barbara, CA 93106-9530, USA}

\author[0000-0002-6790-2397]{Zhou Fan}
\affiliation{National Astronomical Observatories, Chinese Academy of Sciences, Beijing 100101, People's Republic of China}
\affiliation{School of Astronomy and Space Science, University of Chinese Academy of Sciences, Chinese Academy of Sciences, Beijing 100049, People's Republic of China}

\author{Estefania Padilla Gonzalez}
\affiliation{Space Telescope Science Institute 3700 San Martin Drive, Baltimore, MD 21218, USA}

\author[0000-0003-4253-656X]{D. Andrew Howell}
\affiliation{Las Cumbres Observatory, 6740 Cortona Drive, Suite 102, Goleta, CA 93117-5575, USA}
\affiliation{Department of Physics, University of California, Santa Barbara, CA 93106-9530, USA}

\author[0000-0002-8028-0991]{Dieter Hartmann}
\affiliation{Department of Physics and Astronomy, Clemson University, Clemson, SC29631, USA}

\author{Jing-Wei Hu}
\affiliation{National Astronomical Observatories, Chinese Academy of Sciences, Beijing 100101, People's Republic of China}

\author[0000-0002-9404-5650]{P\'all Jakobsson}
\affiliation{Center for Astrophysics and Cosmology, Science Institute, University of Iceland, Dunhagi 5, 107 Reykjavík, Iceland}

\author[0000-0001-5798-4491]{Cheng-Kui Li}
\affiliation{Key Laboratory of Particle Astrophysics, Institute of High Energy Physics, Chinese Academy of Sciences, Beijing 100049, People’s Republic of China}

\author{Zhi-Xing Ling}
\affiliation{National Astronomical Observatories, Chinese Academy of Sciences, Beijing 100101, People's Republic of China}
\affiliation{School of Astronomy and Space Science, University of Chinese Academy of Sciences, Chinese Academy of Sciences, Beijing 100049, People's Republic of China}

\author[0000-0001-5807-7893]{Curtis McCully}
\affiliation{Las Cumbres Observatory, 6740 Cortona Drive, Suite 102, Goleta, CA 93117-5575, USA}
\affiliation{Department of Physics, University of California, Santa Barbara, CA 93106-9530, USA}

\author[0000-0001-9570-0584]{Megan Newsome}
\affiliation{Las Cumbres Observatory, 6740 Cortona Drive, Suite 102, Goleta, CA 93117-5575, USA}
\affiliation{Department of Physics, University of California, Santa Barbara, CA 93106-9530, USA}

\author[0000-0003-4876-7756]{Benjamin Schneider}
\affiliation{Massachusetts Institute of Technology, Kavli Institute for Astrophysics and Space Research, Cambridge, Massachusetts, United States}
\affiliation{Aix Marseille Université, CNRS, CNES, LAM, Marseille, France}

\author{Kaew Samaporn Tinyanont}
\affiliation{National Astronomical Research Institute of Thailand, 260 Moo 4, T. Donkaew, A. Maerim, Chiangmai, 50180 Thailand}

\author{Ning-Chen Sun}
\affiliation{National Astronomical Observatories, Chinese Academy of Sciences, Beijing 100101, People's Republic of China}
\affiliation{School of Astronomy and Space Science, University of Chinese Academy of Sciences, Chinese Academy of Sciences, Beijing 100049, People's Republic of China}
\affiliation{Institute for Frontier in Astronomy and Astrophysics, Beijing Normal University, Beijing 102206, People's Republic of China}

\author[0000-0003-0794-5982]{Giacomo Terreran}
\affiliation{Adler Planetarium 1300 S Dusable Lk Shr Dr, Chicago, IL 60605, USA}

\author[0000-0001-7471-8451]{Qing-Wen Tang}
\affiliation{Department of Physics, School of Physics and Materials Science, Nanchang University, Nanchang 330031, People's Republic of China}

\author{Wen-Xin Wang}
\affiliation{National Astronomical Observatories, Chinese Academy of Sciences, Beijing 100101, People's Republic of China}

\author{Jing-Jing Xu}
\affiliation{Key Laboratory of Particle Astrophysics, Institute of High Energy Physics, Chinese Academy of Sciences, Beijing 100049, People’s Republic of China}

\author{Wei-Min Yuan}
\affiliation{National Astronomical Observatories, Chinese Academy of Sciences, Beijing 100101, People's Republic of China}
\affiliation{School of Astronomy and Space Science, University of Chinese Academy of Sciences, Chinese Academy of Sciences, Beijing 100049, People's Republic of China}

\author[0000-0003-4111-5958]{Bin-Bin Zhang}
\affiliation{School of Astronomy and Space Science, Nanjing University, Nanjing 210093, People's Republic of China}
\affiliation{Purple Mountain Observatory, Chinese Academy of Sciences, Nanjing 210023, People's Republic of China}
\affiliation{Key Laboratory of Modern Astronomy and Astrophysics (Nanjing University), Ministry of Education, People's Republic of China}

\author{Hai-Sheng Zhao}
\affiliation{Key Laboratory of Particle Astrophysics, Institute of High Energy Physics, Chinese Academy of Sciences, Beijing 100049, People’s Republic of China}

\author[0000-0001-8869-0672]{Juan Zhang}
\affiliation{Key Laboratory of Particle Astrophysics, Institute of High Energy Physics, Chinese Academy of Sciences, Beijing 100049, People’s Republic of China}

\begin{abstract}
    We present multiband observations and analysis of EP240801a, a low-energy, extremely soft gamma-ray burst (GRB) discovered on August 1, 2024 by the Einstein Probe (EP) satellite, with a weak contemporaneous signal also detected by Fermi/GBM. Optical spectroscopy of the afterglow, obtained by GTC and Keck, identified the redshift of $z = 1.6734$. EP240801a exhibits a burst duration of 148~s in X-rays and 22.3~s in gamma-rays, with X-rays leading by 80.61~s. Spectral lag analysis indicates the gamma-ray signal arrived 8.3 s earlier than the X-rays. Joint spectral fitting of EP/WXT and Fermi/GBM data yields an isotropic energy $E_{\gamma,\rm{iso}} = (5.57^{+0.54}_{-0.50})\times 10^{51}\,\rm{erg}$, a peak energy $E_{\rm{peak}} = 14.90^{+7.08}_{-4.71}\,\rm{keV}$, a ﬂuence ratio S$(25-50\,\rm{keV})/$S$(50-100\,\rm{keV}) = 1.67^{+0.74}_{-0.46}$, classifying EP240801a as an X-ray flash (XRF). The host-galaxy continuum spectrum, inferred using \texttt{Prospector}, was used to correct its contribution for the observed outburst optical data. Unusual early $R$-band behavior and EP/FXT observations suggest multiple components in the afterglow. Three models are considered: two-component jet model, forward-reverse shock model and forward-shock model with energy injection. Both three provide resonable explanations. The two-component jet model and the energy injection model imply a relatively small initial energy and velocity of the jet in the line of sight, while the forward-reverse shock model remains typical. Under the two-component jet model, EP240801a may resemble GRB 221009A (BOAT) if the bright narrow beam is viewed on-axis. Therefore, EP240801a can be interpreted as an off-beam (narrow) jet or an intrinsically weak GRB jet. Our findings provide crucial clues for uncovering the origin of XRFs.

\end{abstract}

\keywords{Gamma-ray bursts (629), X-ray transient sources (1852), Transient sources (1851)}

\section{Introduction} \label{sec:intro}
Gamma-ray bursts (GRBs) are extraordinarily energetic and luminous catastrophic events in the universe, with a typical isotropic energy $10^{50}-10^{55}\, \rm{erg}$ released in the prompt emission phase \citep{intro1_Atteia}. The duration of GRBs ($T_{90}$, defined as the time span from 5$\%$ to 95$\%$ of the total prompt emission fluence) typically ranges from subseconds to several thousand seconds \citep{intro2_K&Z}. The statistics of the duration of the GRB prompt emission $T_{90}$ reveal two primary types of GRB \citep{T90diff_1,T90diff_2}: Type I bursts (most with $T_{90}$ $<$ 2 s, so they are also called ``short GRBs") and Type II bursts (most with $T_{90}$ $>$ 2 s, known as ``long GRBs"). From the 1234 BATSE GRB samples\footnote{\url{https://gammaray.msfc.nasa.gov/batse/grb/catalog/4b/index.html}} \citep{BATSE4B}, Type II bursts constitute about 70$\%$ of the total GRB sample, originate from the collapse of massive stars, and are associated with broad-lined Type Ic supernovae \citep{Galama99,Woosley06,intro2_K&Z}. In contrast, about 30$\%$ of the GRBs are classified as Type I bursts, which originate from compact star mergers and are accompanied by kilonovae \citep{T90diff_1,T90diff_2,Abbott17}. However, there are some exceptions, e.g., long duration Type I and short duration Type II GRBs such as GRB 060505 \citep{Fynbo06}, GRB 060614 \citep{Della06,Fynbo06,Gehrels06,2015NatCo...6.7323Y}, GRB 200826A \citep{Ahumada01,Zhang21}, GRB 211221A \citep{1211ARastinejad,1211ATroja}, and GRB 230307A \citep{230307ALevan,230307AYang}.

As extensions of the classical GRBs, there are some fainter events, e.g., low-luminosity GRBs (i.e., \citealt{GRB980425};   \citealt{GRB031203}; \citealt{GRB060218}), X-ray rich GRBs (see \citealt{Sakamoto_2008}) and X-ray flashes (see \citealt{HETEXRF}), whose physical origins remain a mystery. Possible scenarios include off-axis GRBs, structured jets, dirty jets, shock breakout of a mildly relativistic jet.

The Einstein Probe (EP), launched on January 9, 2024, is dedicated to monitoring the soft X-ray sky \citep{Yuan2025}. The satellite is equipped with a Wide-field X-ray Telescope (WXT; 0.5–-4 keV, \citealt{EPhandbook}) and a Follow-up X-ray Telescope (FXT; 0.3–-10 keV, \citealt{FXT}). The WXT's lobster-eye micropore optics provide an expansive field of view of $\sim 3600$ square degrees with a sensitivity of $\sim 2.6 \times 10^{-11} \rm erg\,s^{-1}\,cm^{-2}$ in 0.5–-4 keV with 1 ks exposure, which is a significant advantage for detecting transients in space. The autonomous follow-up observations by FXT provide quick positions accurate to $\sim 10''$. Several intriguing transients have already been detected, such as EP240219a \citep{ep240129a}, EP240315a \citep{2024arXiv240416350L, Gillanders2024, ep240315a}, EP240408a \citep{ep240408a}, and EP240414a \citep{ep240414a, 2024arXiv240919056V, Srivastav2025, bright2024_ep240414a}. 
Thanks to the high sensitivity and wide-field monitoring in soft X-rays, EP will be efficient in detecting weak GRBs, providing a great opportunity to unveil the physics behind.

\begin{figure*}[htbp!]
\includegraphics[width=1.0\textwidth, keepaspectratio]{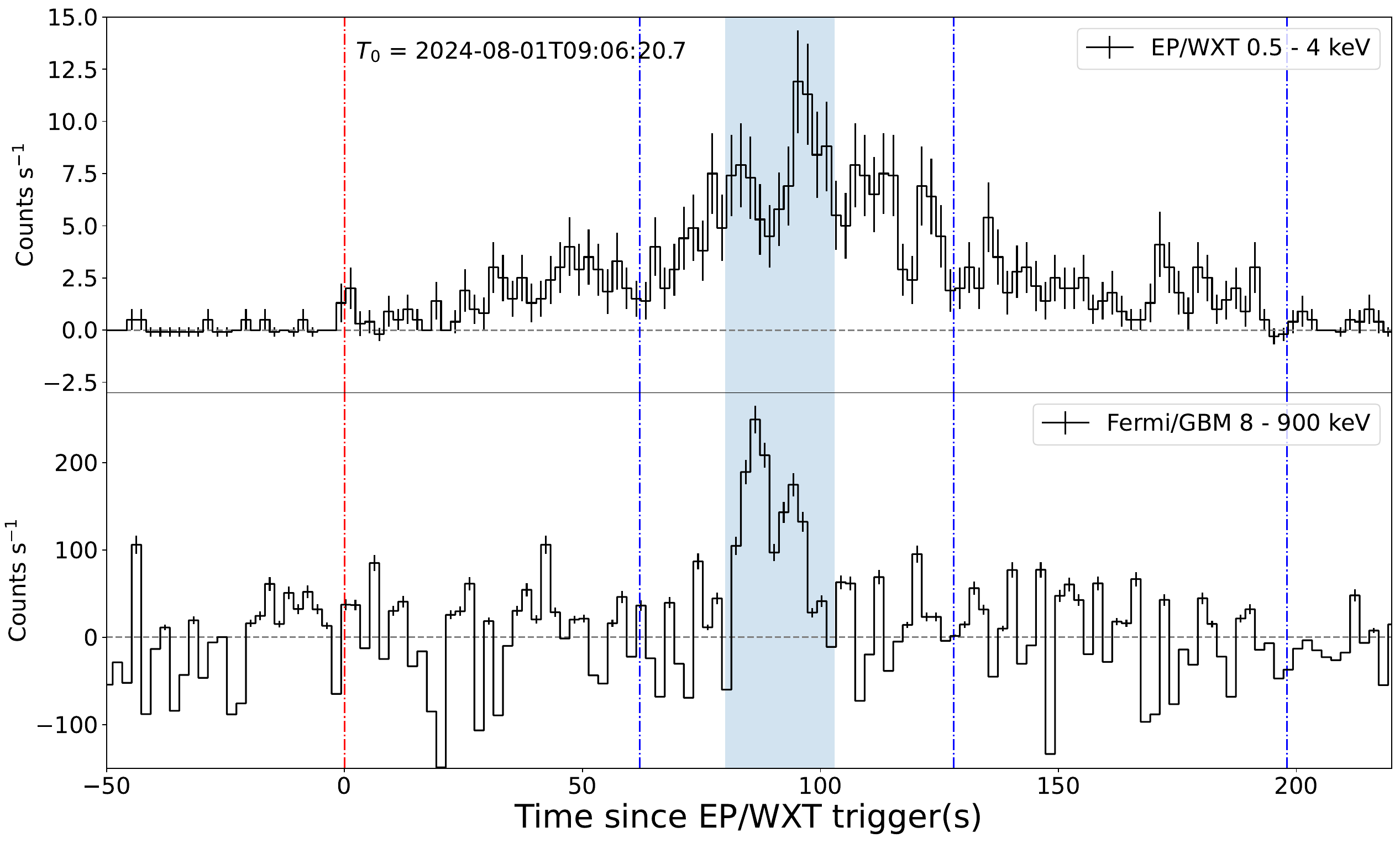}
\caption{The prompt emission light curve obtained by EP/WXT and Fermi/GBM, the grey dotted lines indicate zero count rate. The red dash-dotted line represents $T_0$ of EP240801a, and the UTC time is indicated. We analyzed the spectrum of Fermi/GBM data during the vertical light-blue slice. The spectral evolution analysis of EP/WXT data was conducted in the three epochs divided by red and blue dash-dotted lines.
\label{promptlc}}
\end{figure*}

In this work, we present an extremely soft GRB detected by EP/WXT, EP240801a, analyze and discuss the mechanism of the burst in detail. EP/WXT detected EP240801a with an uncertainty of $\sim 3'$, and EP/FXT rapidly performed a follow-up observation, which reduced the positional uncertainty to $\sim 10''$. As a result, the field of EP240801a was observed in multiple bands and a spectroscopic redshift of EP240801a was discovered soon after the trigger \citep{GCN37013,TRTGCN,Zheng24}. The observed data collected from several facilities are introduced in Section \ref{sec:data}. Section \ref{sec:result} presents our analysis of the prompt emission phase, afterglow phase, and the host galaxy. In Sections \ref{sec:discussion} and \ref{sec:summary}, we respectively discuss the models of EP240801a and summarize our work. The conventional cosmological model we adopted is as follows: H$_0 = 69.6\,\rm{km}\,\rm{s}^{-1}\,\rm{Mpc}^{-1},\,\Omega_M = 0.286,\,\Omega_{\Lambda} = 0.714$ \citep{Cosmic_param}.

\begin{figure}[ht!]
\centering
\includegraphics[width=0.9\linewidth]{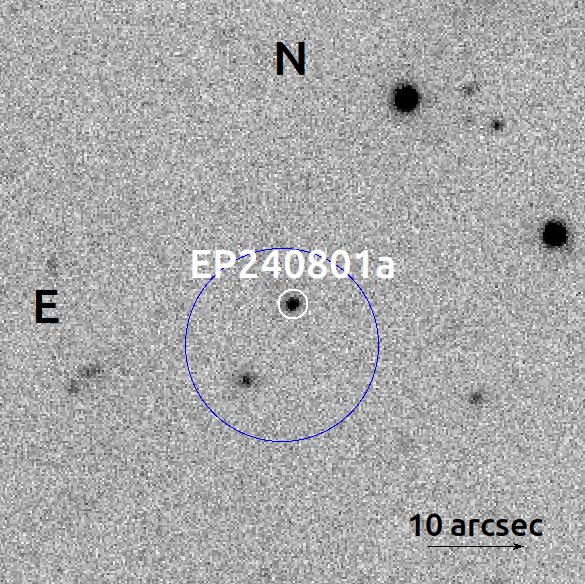}
\caption{The Sloan $r'$-band position of EP240801a with the FOV $1^{'} \times 1^{'}$ obtained by NOT/ALFOSC about 0.8 days after the discovery of EP240801a. The location of the burst is circled in white and the 10$\arcsec$ error circle of EP/FXT is in blue.
\label{locimg}}
\end{figure}

\section{Observations and Data Reduction} \label{sec:data}
\subsection{Einstein Probe Observations}
EP240801a triggered WXT at 09:06:03 on 2024-08-01 (UTC dates are used throughout this paper), and FXT autonomous observations followed $\sim 180$ s later \citep{EP240801aGCN}. The transient source has an unabsorbed peak flux $(1.15_{-0.08}^{+0.06}) \times 10^{-8}\,\rm{erg\,s^{-1}\,cm^{-2}}$ in 0.5–-4~keV, and it is located at R.A.~=~345.1630 deg, Dec.~=~32.5927 deg (J2000) with an uncertainty of 10$\arcsec$ in radius (at 90$\%$ confidence, statistical and systematic). After that, FXT performed six more follow-up observations until 6.5 days after the burst, and EP240801a was identified in the first four of those observations. 

The processing and calibration of WXT photon events are handled adopting specialized data-reduction software and the calibration database (Y. Liu et al., in prep.). The calibration database is generated on the basis of the results of the on-ground calibration experiments (H.-Q. Cheng et al., in prep.). The position of each photon was converted to celestial coordinates (J2000). The energy value of each event is calculated according to the bias and gain stored in the calibration database. After bad/flaring pixels were flagged, single, double, triple and quadruple events without anomalous flags were selected to form the cleaned event file. The photons of the source and the background were extracted from a circle with a radius of $9'$ and an annulus with radii of $18'$ and $36'$, respectively. As the WXT average net count rate is $\sim 2.9$ in the total prompt emission phase, we grouped the WXT data with 5 minimum counts per bin to perform the spectral analysis.

The FXT cleaned event ﬁles and response files were generated by using the Follow-up X-ray Telescope Data Analysis Software (\texttt{FXTDAS v1.10} )\footnote{\url{http://epfxt.ihep.ac.cn/analysis}}.
The process involved particle event identification, pulse invariant conversion, grade calculation and selection (grade $\le$ 12), bad- and hot-pixel flag and selection of good time intervals using housekeeping file. With the $90\%$ of the Point Spread Function (PSF) is enclosed by a $\sim 1'$ radius circle at 1.5 keV, the photons of source and background were extracted from a circle with a radius of $1'$ and an annulus with radii of $2'$ and $3'$, respectively. The FXT data at various times were also grouped with different counts to enhance the signal-to-noise ratio (SNR).

The start time of EP240801a was determined to be 09:06:20.7 on 2024-08-01  (hereafter $T_0$) by using the Bayesian block method \citep{Bayesianblock} after reprocessing the WXT data; this is 17.7 s later than the start time in General Coordinates Network (GCN) circular reported by \citet{EP240801aGCN}.
The prompt emission shows a multipeaked structure with $T_{90,\,0.5-4\, \rm{keV}}=148.0\,\pm\,3.2\,\rm{s}$ as illustrated in Fig.~\ref{promptlc}. Spectral analysis was performed with \texttt{Xspec v12.14.0h} \citep{Xspec} for EP data; details can be seen in Section \ref{sec:high-energy spec}. The results of the EP observations are presented in Table \ref{Xray_result}.

\begin{figure*}[htbp!]
\gridline{\fig{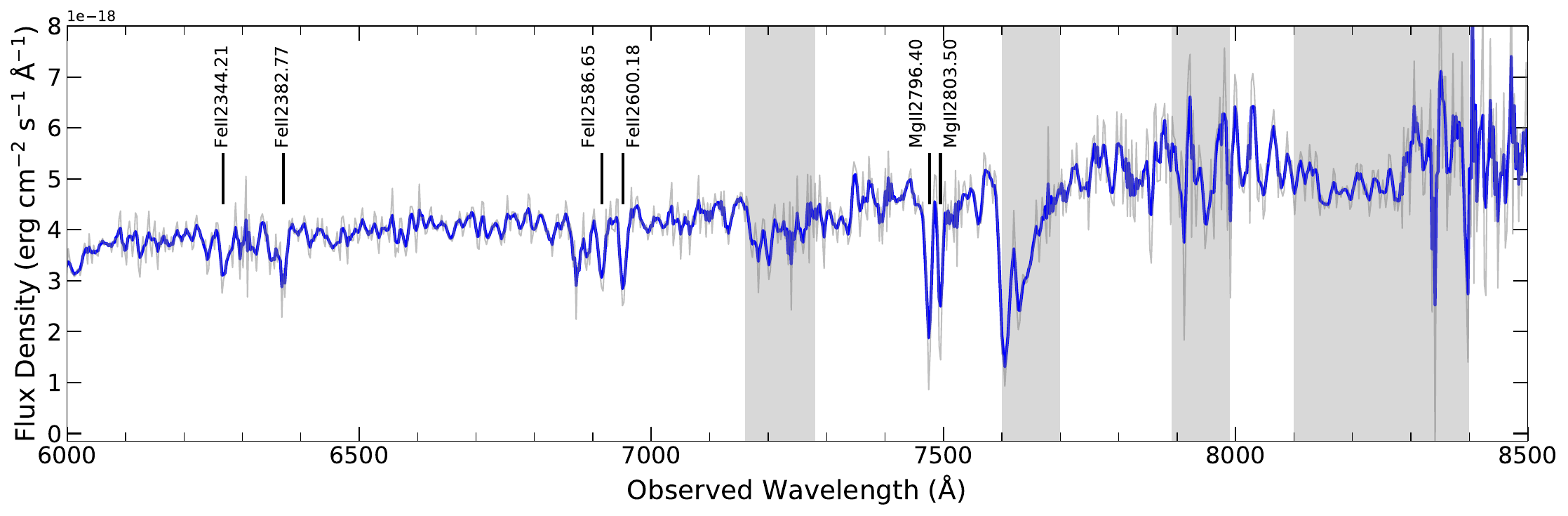}{0.9\textwidth}{}}
\gridline{\fig{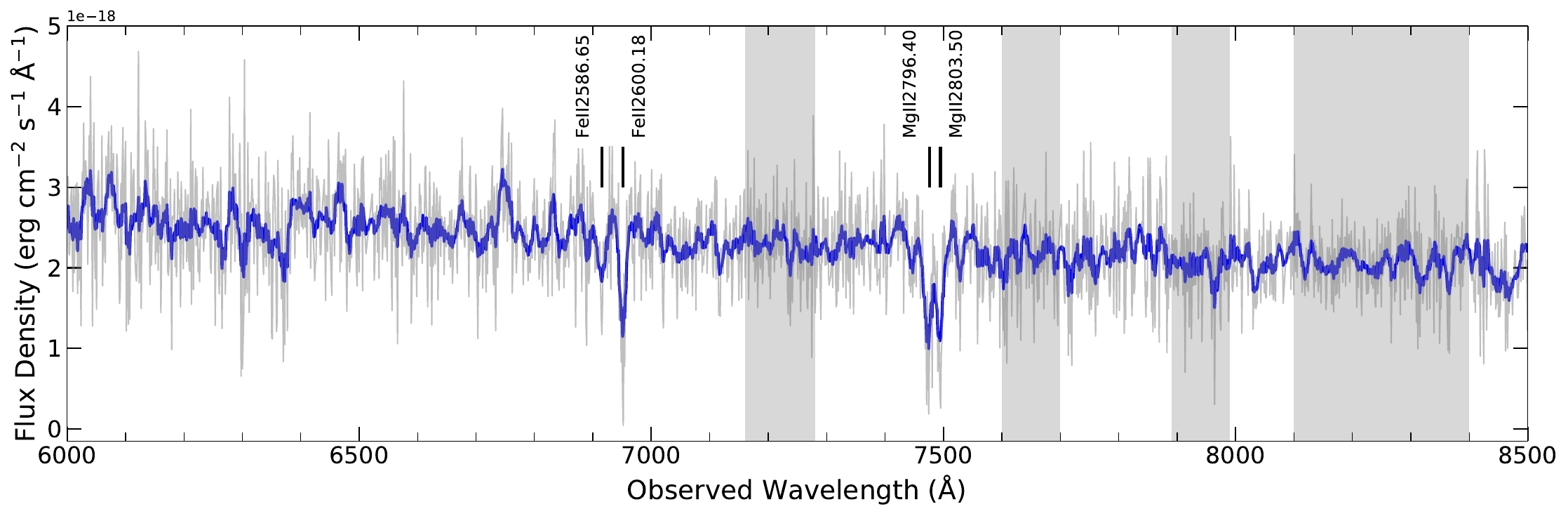}{0.9\textwidth}{}}
\caption{The spectrum obtained with GTC/OSIRIS+ (top panel) and Keck/LRIS (bottom panel). The gray line represents the raw spectrum, while the blue line has been smoothed for display purposes. The identified metal absorption lines are indicated with vertical dashes and the gray vertical bands indicate the locations of telluric features. From the observed absorption lines, especially Mg~II $\lambda\lambda2796$, 2803 and Fe~II $\lambda$2600, we determine $z = 1.6734\pm 0.0002$ for EP240801a, assuming the transient occurred in the host galaxy at that redshift.\label{optspec}}
\end{figure*}

\subsection{Fermi Observations}
Based on the trigger time and location of the burst, we found a weak, subthreshold signal 80.61 s after $T_0$ by using the Bayesian block method in the archive data of the Gamma-ray Burst Monitor (GBM, \citealt{GBM}) onboard the Fermi Gamma-Ray Space Telescope (Fermi). This signal was detected in the four closest detectors (n6, n7, n9, and nb) of GBM with a duration $T_{90,\,8-900\,\rm{keV}}$ of $22.30\,\pm\,9.92\,\rm{s}$. By fitting a first-order polynomial to the GBM data before and after the signal (i.e., $-100$ to 70 s and 150 to 200 s referenced to $T_0$, respectively), we eliminated the background contribution, providing the background-subtracted  gamma-ray light curve, as depicted in Fig.~\ref{promptlc}.

Spectral analysis was conducted using the four closest sodium iodide (NaI; 8 keV--1 MeV) detectors. The two bismuth germanate (BGO; 200 keV--40 MeV) detectors were excluded from the spectral analysis due to the absence of significant transient signals in their data. We excluded the GBM data between 30 keV and 40 keV (corresponding to the iodine K-edge; see \citealp{GBM}) and the channels at the extremes of the spectra (channels below 8 keV and channels 127 and 128 for NaI). We obtained the time-tagged event data covering the time range of EP240801a from the Fermi/GBM public data archive\footnote{\url{https://heasarc.gsfc.nasa.gov/FTP/fermi/data/gbm/daily/}}. \texttt{The Multi-Mission Maximum Likelihood Framework} (\texttt{threeML}; \citealt{3ml}) is the main tool for analysis of Fermi/GBM data.

\subsection{Ground-Based Observations}
The optical counterpart of EP240801a was first detected by the 0.7~m telescope of the Thai Robotic Telescope (TRT) network at 11:06:18 on 2024-08-01, 2.24 hr after the EP/WXT trigger. It was located at (J2000) R.A.~=~$23^{\rm hr}00^{\rm m}39.03^{\rm s}$, Dec.~=~$+32^\circ 35' 37.95''$ with an uncertainty of $0.5''$; this is $4.4''$ away from the FXT center position \citep{TRTGCN}. The celestial location of the burst is shown in Fig.~\ref{locimg}.

The telescopes that we used for our multiband photometric follow-up observations are listed in the Appendix~\ref{sec:photometic table}.  

After standard data reduction with \texttt{IRAF} \citep{IRAF}, the HiPERCAM pipeline and the pyEMIR pipeline, and astrometric calibration by Astrometry.net \citep{solve-field}. The optical photometry was calibrated with the Legacy Surveys Data Release 10 \citep{legacy} and the Sloan Digital Sky Surveys Data Release 18 \citep{SDSS}, while the near-infrared (NIR) data were calibrated with the Two Micron All Sky Survey (2MASS) catalog \citep{2MASS}. The photometry in Johnson–Cousin filters were calibrated with the converted magnitude from the Sloan system\footnote{\url{https://live-sdss4org-dr12.pantheonsite.io/algorithms/sdssUBVRITransform/\#Lupton}} for the nearby reference stars. 
It should be noted that there may be an additional systematic error for the differential photometry of $u',\,H,\,Ks$ bands, as the catalogues are significantly shallower than the target.
Details of the filters used with these telescopes and the photometric results are presented in Table \ref{tab:optical_result} and shown in Fig.~\ref{totallc}.

\subsection{Optical Spectroscopy}

The optical spectra of EP240801a were obtained with the Optical System for Imaging and low-intermediate-Resolution Integrated Spectroscopy (OSIRIS+) instrument mounted on the 10.4~m Gran Telescopio Canarias (GTC, \citealp{cepa_2000}) at $\sim 0.79$ days after the EP/WXT trigger time, with an exposure time of $4 \times 1200$ s \citep{GCN37013} and the Low Resolution Imaging Spectrometer (LRIS) mounted on Keck telesocpe \citep{Oke_1995} at $\sim 1.02$ days, with an exposure time of $3 \times 950$~s \citep{Zheng24}.

The OSIRIS+ observations utilized the 1"-wide slit oriented along the parallactic angle and the R1000R grism, which has a coverage of 5100~-~10,000\,\r{A}. The data reduction followed bias subtraction and flat-field correction using the standard \texttt{PyRAF} tasks \citep{pyraf2012} and cosmic rays correction with the \texttt{LACosmic} task \citep{vanDokkum2001}.

\begin{deluxetable*}{cccccc}
\tablecaption{Spectral Fitting Results and Corresponding Fitting Statistics for EP and Fermi/GBM
\label{Xray_result}}
\tablewidth{0pt}
\tablehead{
\colhead{Instruments} & \colhead{Time Intervals} & \colhead{Intrinsic Absorption} & \colhead{Photon Index$^{*}$} & \colhead{$E_{\rm{peak}}$} & \colhead{STAT/(d.o.f.)$^{\dag}$}\\ 
\colhead{} & \colhead{(second)}  & \colhead{(cm$^{-2}$)} & \colhead{($\Gamma$)} & \colhead{(keV)} &  \colhead{}
}
\startdata
    \multirow{4}{*}{WXT} & 0 - 198 & $(2.13 \pm 0.54) \times 10^{22}$ & $1.99 \pm 0.18$ & -- & 79.73/83\\
    & 0 - 62 & $\cdots$ & $1.91 \pm 0.30$ & -- & 8.84/14\\
    & 62 - 128 & $\cdots$ & $1.56 \pm 0.14$ & -- & 46.09/47\\
    & 128 - 198 & $\cdots$ & $2.53 \pm 0.20$ & -- & 37.21/34\\
    GBM & 80 - 103 & -- & $1.37 \pm 0.36$ & $18.03^{+9.93}_{-5.51}$ & 155.76/156 \\
    WXT+GBM &  $\cdots$ & $\cdots$ & $1.65 \pm 0.08$ & $14.90^{+7.08}_{-4.71}$ & (46.70+154.40)/205 \\
    \multirow{3}{*}{FXT} & 232 - 697 & $<0.07\times10^{22}$ & $3.29 \pm 0.07$ & -- & 62.13/53 \\
    & 3.75 $\times 10^3$ - 6.30 $\times 10^3$ & $(0.56 \pm 0.45) \times10^{22}$ & $2.18 \pm 0.15$ & -- & 24.79/33 \\
    & 1.54 $\times 10^5$ - 1.41 $\times 10^6$ & $<1.11\times10^{22}$ & $1.73_{-0.15}^{+0.28}$ & -- & 59.77/49 \\
\enddata
\tablecomments{
\begin{itemize}
\item[] Ellipses ($\cdots$) in Time Intervals and Intrinsic Absorption indicate that we use the same value as the result above.
\item[] All error bars represent $1\sigma$ uncertainties. 
\item[*]  An absorbed power-law model ($zTbAbs\times TbAbs\times PowerLaw$) is used to fit the X-ray data, and the Galactic hydrogen column density is fixed with $N_{\rm{H}} = 9.78 \times 10^{20}\,{\rm cm}^{-2}$. When GBM data are involved, $PowerLaw$ is replaced by $Cutoff-PowerLaw$.
\item[\dag] We use PGSTAT for GBM data and CSTAT for EP/WXT and EP/FXT data.
\end{itemize}
}
\end{deluxetable*}

For Keck/LRIS, the spectrum was acquired with the slit oriented near the parallactic angle to minimize slit losses caused by atmospheric dispersion \citep{Filippenko_1982}. The LRIS observations utilized the $1\arcsec$-wide slit, 600/4000 grism, and 400/8500 grating,
which produced a spectral coverage of 3140--10,270\,\r{A}.
Data reduction followed standard techniques for CCD processing and spectrum extraction using the LPipe data-reduction pipeline \citep{Perley_2019}. Low-order polynomial fits to comparison-lamp spectra were used to calibrate the wavelength scale, and small adjustments derived from night-sky lines in the target frames were applied. The spectrum was flux calibrated using observations of appropriate spectrophotometric standard stars observed on the same night, at similar airmasses, and with an identical instrument configuration; these standard-star spectra were also used to remove telluric absorption.

The sections of the spectra with the highest SNR (6000--8500\,\r{A}) are illustrated in Fig.~\ref{optspec}. The significant Mg~II doublet and some Fe~II absorption lines lead to a redshift of $z = 1.6734\pm 0.0002$. Hence we consider this value as the redshift of EP240801a, based on the reasonable assumption that the transient occurred in the galaxy producing those lines rather than one at a greater distance.

\section{Results} \label{sec:result}

\subsection{Prompt Emission}
\subsubsection{Light Curve}\label{sec:Promptlc}
In Fig.~\ref{promptlc}, we present the prompt emission light curves of EP240801a with EP/WXT data at 0.5--4 keV and Fermi/GBM data at 8--900 keV, both with 2~s bin size. The prompt emission shows  multiple pulses at 0.5--4 keV with $T_{90} = 148.0\,\pm\,3.2\,\rm{s}$. Only two peaks within $T_{90} = 22.30\,\pm\,9.92\,\rm{s}$ are visible at 8--900 keV in the light curve from Fermi/GBM data. The spectral lag of EP/WXT and Fermi/GBM data was calculated by using the cross-correlation function (CCF, \citealp{Band97,Norris_2000,U10}) method with 500~ms time bin. For the uncertainty of lags, we used the Monte Carlo simulation (see \citealt{Peterson98,U10}). The lag value is $8.3 \pm 0.8$ seconds, which indicates the gamma-ray signal arrived 8.3 s earlier.

\begin{figure}[ht!]
\includegraphics[width=0.45\textwidth, keepaspectratio]{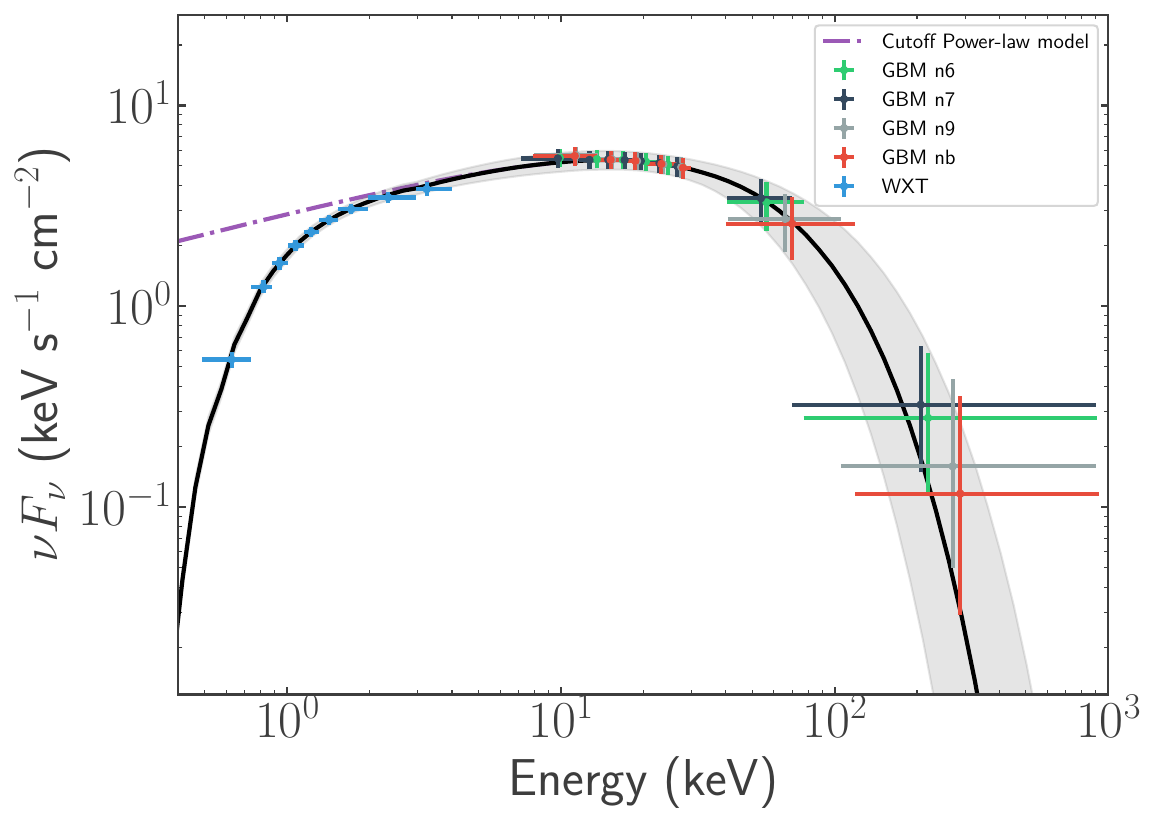}
\caption{The spectral joint fitting of EP/WXT and Fermi/GBM data during the light-blue slice in Fig.~\ref{promptlc}. The black solid line represents the best-fit absorbed cutoff power-law model and the shadow is the $1\sigma$ uncertainty interval, while the violet dash-dotted line shows the individual cutoff power-law model. The data from different detectors are marked as various colors.
\label{joint fit}}
\end{figure}

\subsubsection{Analysis of the high-energy spectra} \label{sec:high-energy spec}

\begin{figure}[ht!]
\includegraphics[width=0.45\textwidth, keepaspectratio]{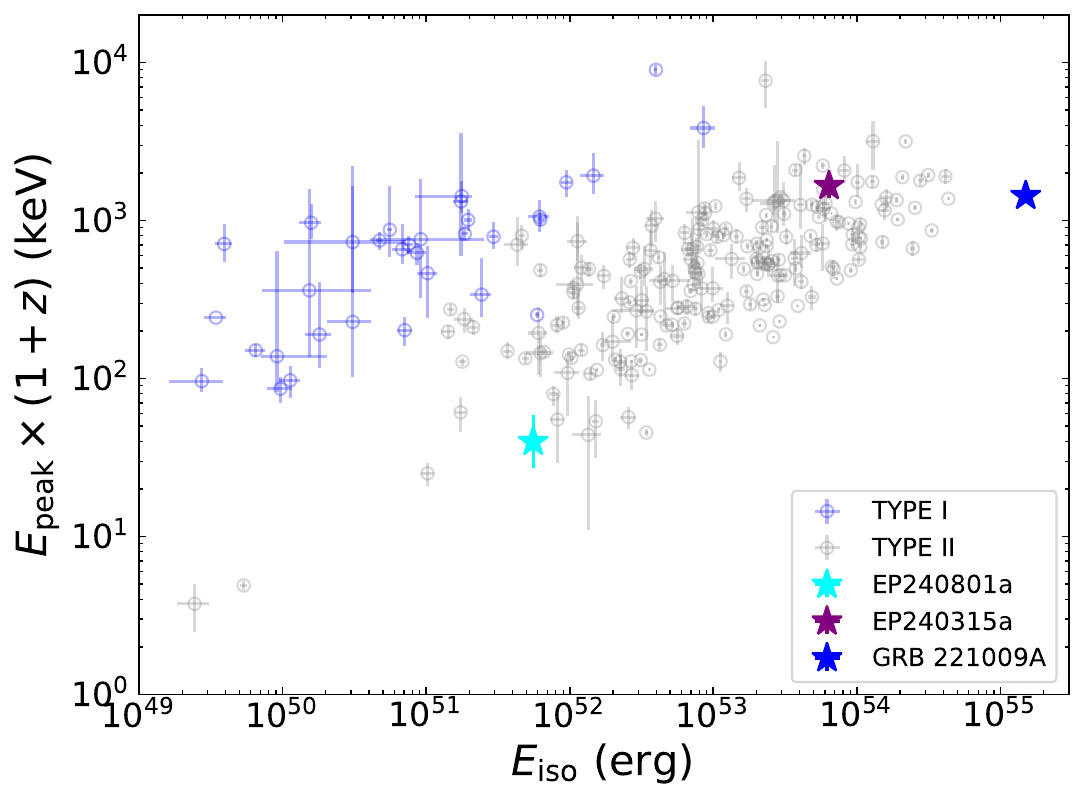}
\caption{Amati diagram of EP240801a (cyan star), the blue and grey circles represent other Type I and Type II GRBs, respectively. Another GRB detected by EP,  EP240315a, is marked as a purple star. GRB 221009A is marked as a blue star.
\label{amatiimg}}
\end{figure}

\begin{figure}[ht!]
\includegraphics[width=0.45\textwidth, keepaspectratio]{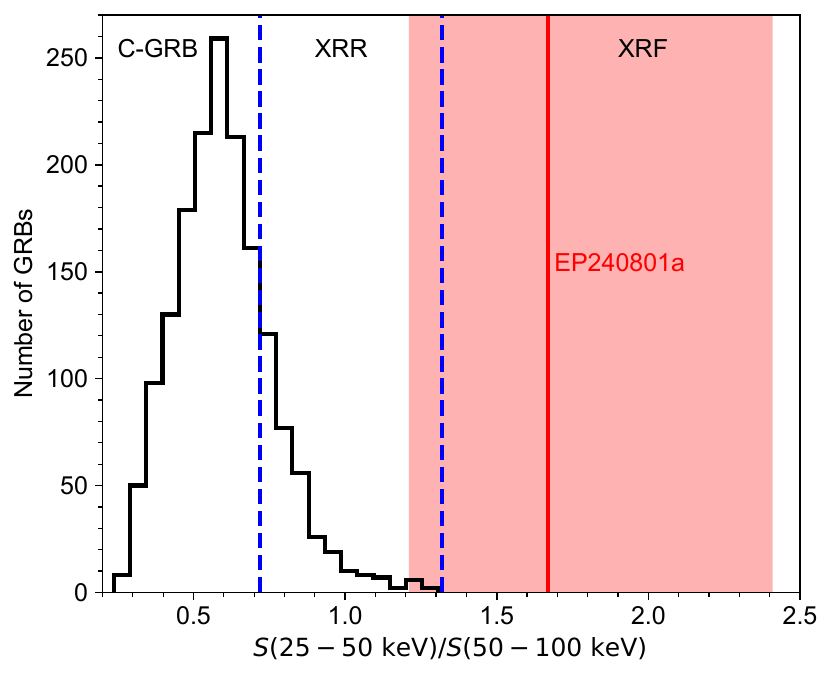}
\caption{The fluence ratio $\rm{S}(25-50\,\rm{keV})/\rm{S}(50-100\,\rm{keV})$ for 1647 GRBs detected by Fermi/GBM during the years 2008--2020. The red line represents the fluence ratio of EP240801a and the shadow is the $1\sigma$ uncertainty interval. The blue dashed lines represent the boundaries of classical GRBs (C-GRBs), X-ray-rich GRBs (XRRs) and X-ray flashes (XRFs) described by \citet{Sakamoto_2008}.
\label{HRimg}}
\end{figure}

\begin{figure*}[ht!]
\includegraphics[width=1.0\textwidth, keepaspectratio]{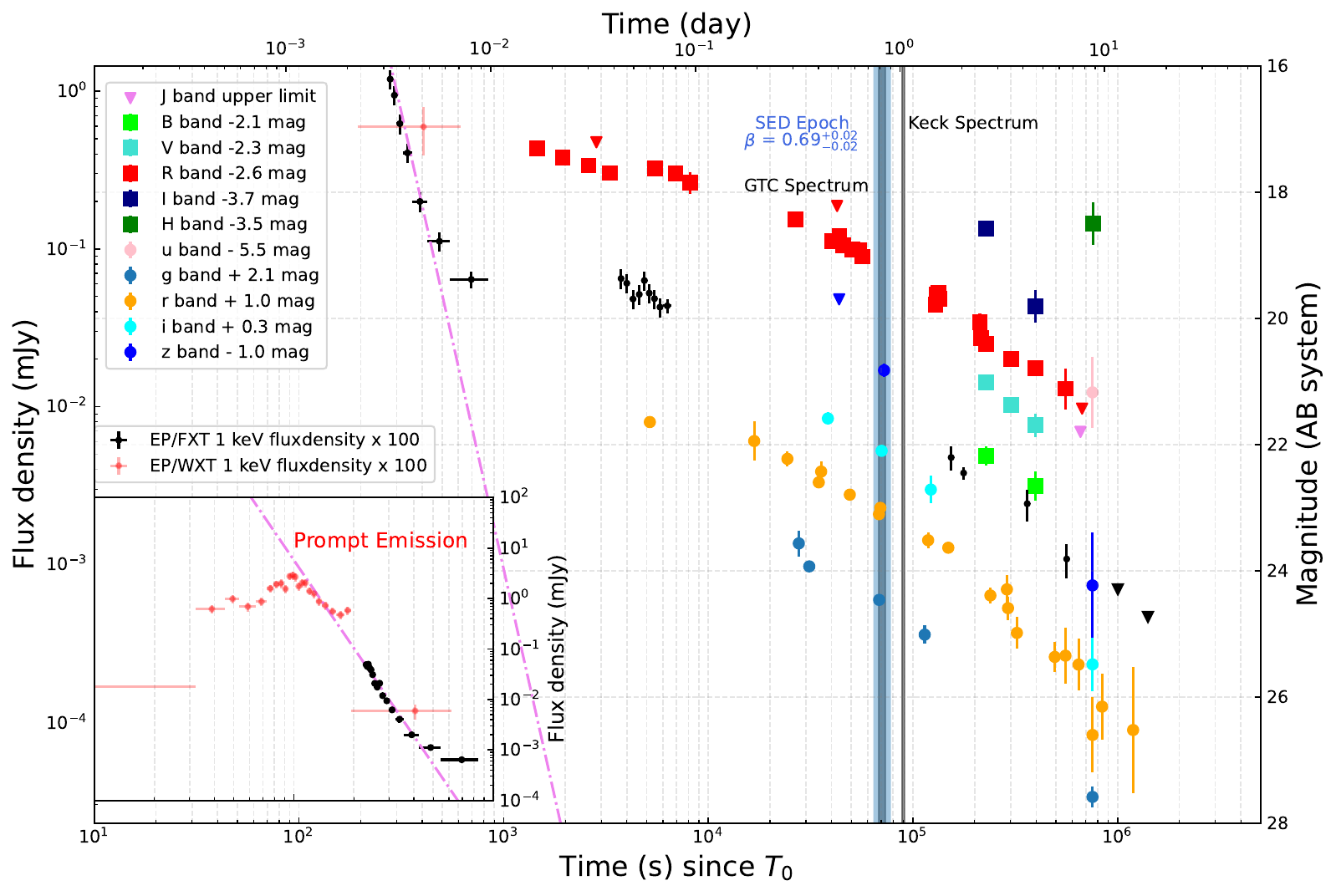}
\caption{Light curves of EP240801a from the prompt phase to the afterglow phase in X-rays, optical, and NIR. The EP/WXT (red circles) and EP/FXT (black circles) data are presented with unabsorbed flux density at 1 keV and scaled by a factor of 100 in the main figure but not scaled in the inset. The inset shares the time axis with the main figure. The optical/NIR data are in the AB system \citep{ABsystem} and have been corrected for Galactic extinction, which is $E(B - V) = 0.093$ mag \citep{SF2011}, and the host-galaxy magnitude. All the inverted triangle symbols signify limits. The light-blue slice represents the time of SED. Two gray slices represent the time of the GTC and the Keck optical spectrum, respectively. A violet dash-dotted single power law with $ \alpha = 5.82\,\pm\,0.32$ is depicted in both the main figure and the inset.
\label{totallc}}
\end{figure*}

We perform X-ray data analyses via \texttt{Xspec} and employ \texttt{threeML} for gamma-ray data and joint fitting. The absorbed power-law model $zTbAbs\times TbAbs\times PowerLaw$ was used to fit the X-ray spectrum. The first and second components are responsible for the intrinsic absorption and the Galactic absorption using the Tuebingen-Boulder interstellar medium absorption model \citep{TbAbs_model}. For the Galactic hydrogen column density, we adopted $N_{\rm{H}} = 9.78 \times 10^{20}\,\rm{cm}^{-2}$, as calculated by the UK Swift Science Data Centre\footnote{\url{https://www.swift.ac.uk/analysis/nhtot/}} using the method described by \cite{NH_method}. The third component is a simple photon power law, with $N(E)=K(\frac{E}{1\,\rm{keV}})^{-\Gamma}$, where $K$ is the normalization of the spectrum and $\Gamma$ is the dimensionless photon index of power law. The cutoff power-law model proves to be more suitable for the GBM spectrum and the X-ray/gamma-ray joint fit of EP240801a, as its results have lower Bayesian information criterion (BIC, \citealt{BIC,BIC2}) value compared to the Band models'. Thus, we used the cutoff power-law model to analyze the GBM data during the light-blue slice in Fig.~\ref{promptlc}. The cutoff power-law model was also utilized to analyze both EP/WXT and GBM data during this period, assuming the Galactic absorption and intrinsic absorption as derived from the EP/WXT analysis before, see Fig.~\ref{joint fit}. The fitting results and the corresponding statistics for each time interval are listed in Table \ref{Xray_result}. We used CSTAT \citep{CSTAT} for EP data and PGSTAT \citep{Xspec} for Fermi/GBM data. The fitting results in differential phase were used to obtain the X-ray flux density at 1 keV, as shown in Fig.~\ref{totallc}. The spectral evolution of EP/WXT data indicates that the spectrum becomes the hardest between 62 and 128 seconds after $T_0$. This could potentially explain why Fermi/GBM detected EP240801a solely during this epoch. We find $E_{\gamma,\rm{iso}} = (5.57^{+0.54}_{-0.50})\times 10^{51}\,\rm{erg}$ in 1--$10^4$ keV in the rest frame, $E_{\rm{peak}} = 14.90^{+7.08}_{-4.71}\,\rm{keV}$, and the ﬂuence ratio $\rm{S}(25-50\,\rm{keV})/\rm{S}(50-100\,\rm{keV}) =  1.67^{+0.74}_{-0.46}$ for EP240801a. The parameters above are shown in Fig.~\ref{amatiimg} and Fig.~\ref{HRimg}. As described by \citet{Sakamoto_2008}, the empirical definitions of X-ray flashes (XRFs), X-ray-rich GRBs (XRRs), and classical GRBs (C-GRBs) are as follows:
\begin{align*}
           & \rm{S}(25-50\,\rm{keV})/\rm{S}(50-100\,\rm{keV})\leq 0.72\,&(\rm C-GRBs),   \\
    0.72 < & \rm{S}(25-50\,\rm{keV})/\rm{S}(50-100\,\rm{keV})\leq 1.32\,&(\rm XRRs),   \\
    1.32 < & \rm{S}(25-50\,\rm{keV})/\rm{S}(50-100\,\rm{keV})\,&(\rm XRFs).
\end{align*}
Thus, EP240801a is classified as an XRF. Despite this classification as X-ray flash, some $\sim$ 100 keV photons are detected. The statistical properties of EP240801a are consistent with the XRFs in the third Swift Burst Alert Telescope catalog (BAT3), as indicated in \cite{2018ApJ...866...97B}, suggesting a greater likelihood of association with a supernova explosion.

\subsection{Afterglow Emission}

\subsubsection{Light Curve}
The X-ray, NIR, and optical data from the prompt to afterglow phases are illustrated in Fig.~\ref{totallc}. The EP/WXT and EP/FXT data during the prompt phase are shown in the inset. The optical and NIR light curves  obtained from $\sim 0.02$ days to $\sim 13.78$ days include $u',\, g',\, r',\, i',\, z',\, B,\, V,\, R,\, I,\, J$, and $H$ bands. During the afterglow phase, the standard external forward shock model predicts a typical decay slope of $\alpha \sim 1$ with a single power-law model $F \propto t^{-\alpha}$ \citep{Zhang2018book}. The X-ray data in the afterglow phase can be divided into two parts. The initial dataset comprises a seemingly shallow decay phase occurring between $\sim 3.7$ and 6.3 ks. And then the X-ray data eventually encounters a normal single power-law decay. Meanwhile, the $R$-band data clearly indicate two distinct epochs of the afterglow phase with different temporal decay indices when each part is fit with the single power law. The initial epoch, occurring prior to 0.1 days with a shallow decay index $\alpha = 0.25\,\pm\,0.05$ for X-ray and $R$-band data, may indicate the presence of an additional emission component, namely a structured jet, a reverse shock along with a forward shock or an energy injection situation. The light curve at later times exhibits a normal decay with $\alpha = 0.95\,\pm\,0.03$ by fitting the $r'$-, $R$-bands and X-ray data.

\subsubsection{Afterglow Spectral Energy Distribution Analysis}
\begin{figure}[ht!]
\includegraphics[width=0.5\textwidth, keepaspectratio]{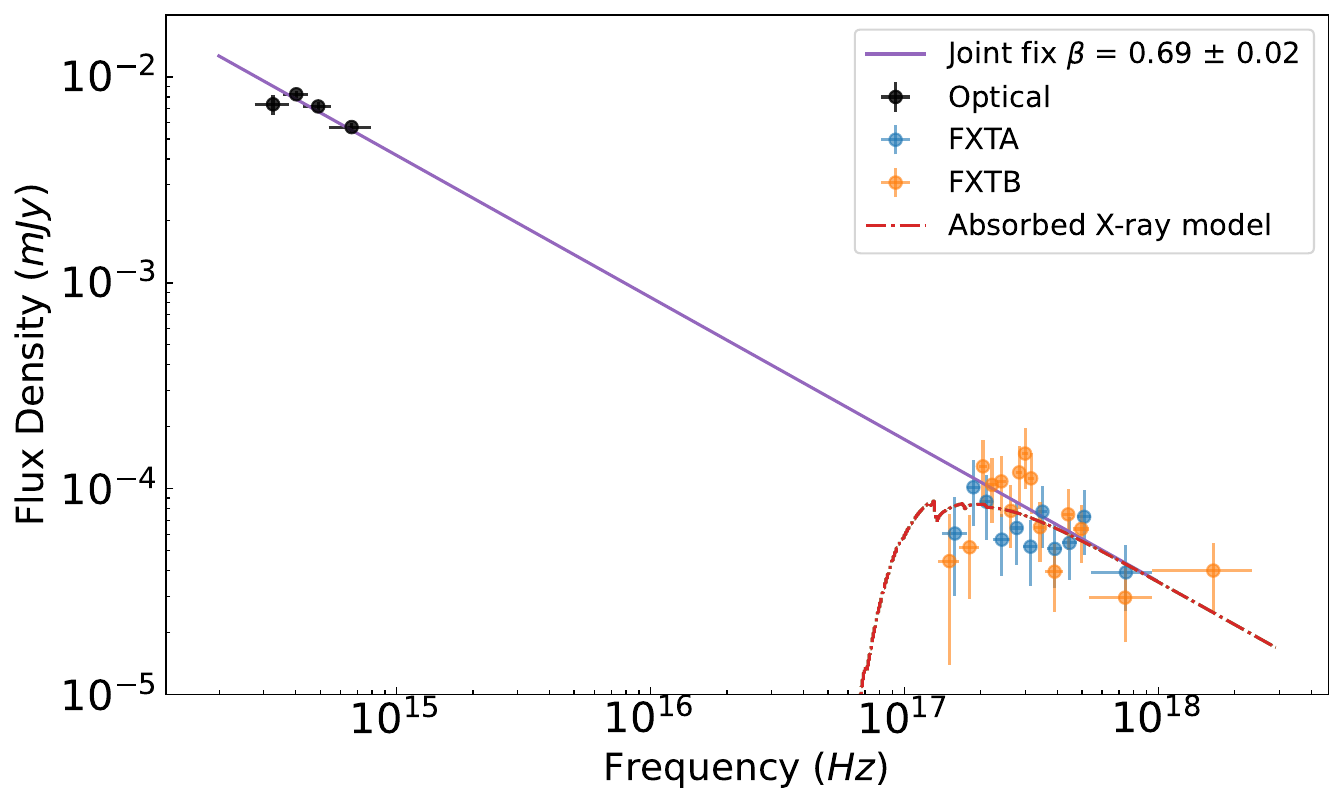}
\caption{The afterglow SED of EP240801a at 70 ks from optical to X-ray frequencies. The optical multiband data are listed in Table \ref{tab:optical_result}. The violet line represents the result of a single power-law fit with $\beta = 0.69 \pm 0.02$. The dash-dotted red line represents the X-ray fit.
\label{sed}}
\end{figure}

A spectral energy distribution (SED) analysis enables a deeper understanding of the afterglow evolution. We conducted an SED analysis at $\sim 70$ ks, denoted by the light-blue region in Fig.~\ref{totallc}. The optical data in the Sloan $g',\, r',\, i',\, z'$ bands obtained from the Alhambra Faint Object Spectrograph and Camera (ALFOSC) mounted on the 2.56~m Nordic Optical Telescope (NOT) and the X-ray data deduced from later times using the temporal decay index $\alpha = 0.95$ were used for the SED; see Fig.~\ref{sed}. We fit the spectrum with a power-law function from the $zDust\times zTbAbs\times TbAbs \times PowerLaw$ model in \texttt{Xspec v12.14.0h}, where $zdust$ represents extinction by dust grains in the host galaxy of the burst, and $zTbAbs$ and $TbAbs$ are, respectively, the intrinsic hydrogen photoelectric absorption in the host galaxy and the Milky Way Galaxy. The redshift and the Galactic hydrogen column density are fixed to 1.6734 and $N_{\rm{H}} = 9.78 \times 10^{20}\,\rm{cm}^{-2}$, respectively. For all three extinction laws (Milky Way, Large Magellanic Cloud, or Small Magellanic Cloud), $E(B - V)$ of the host galaxy cannot be accurately constrained and tends toward zero under the optimal statistical conditions. The best-fitting result gives the spectral index $\beta = 0.69 \pm 0.02$ in the optical to X-ray bands.

\begin{deluxetable}{ccc}
\tablecaption{Host-Galaxy Magnitudes of EP240801a
\label{host_mag}}
\tablewidth{0pt}
\tablehead{
\colhead{Band} & \colhead{Magnitude (AB)} & \colhead{References}
}
\startdata
         $FUV$ & $> $ 20.70 & \multirow{2}{*}{GALEX} \\
         $NUV$ & $> $ 21.40 &  \\
         $u$ & $>$25.48 & GTC \\
         $g$ & 25.13 $\pm$ 0.11 & GTC \\
         $r$ & 24.40 $\pm$ 0.19 & NOT \\
         $i$ & 24.46 $\pm$ 0.20 & NOT \\
         $z$ & $>$ 24.60 & GTC \\
         $J$ & $> $ 21.80 & SAI-25 \\
         $H$ & $> $ 21.52 & GTC \\
         $Ks$ & 22.12 $\pm$ 0.29 & VLT \\
         $W1$ & $> $ 17.74 & \multirow{4}{*}{ALLWISE}\\
         $W2$ & $> $ 16.28 & \\
         $W3$ & $> $ 11.60 & \\
         $W4$ & $> $ 8.63 & \\
         \hline
         $u$ & 25.92 $\pm$ 0.25 & \multirow{7}{*}{\texttt{Prospector} model results} \\
         $z$ & 24.03 $\pm$ 0.25 &  \\
         $B$ & 25.22 $\pm$ 0.25 &  \\
         $V$ & 24.86 $\pm$ 0.25 &  \\
         $R$ & 24.71 $\pm$ 0.25 & \\
         $I$ & 24.27 $\pm$ 0.25 &  \\
         $H$ & 22.64 $\pm$ 0.25 &  \\
\enddata
\end{deluxetable}

\begin{figure}[ht!]
\includegraphics[width=0.5\textwidth, keepaspectratio]{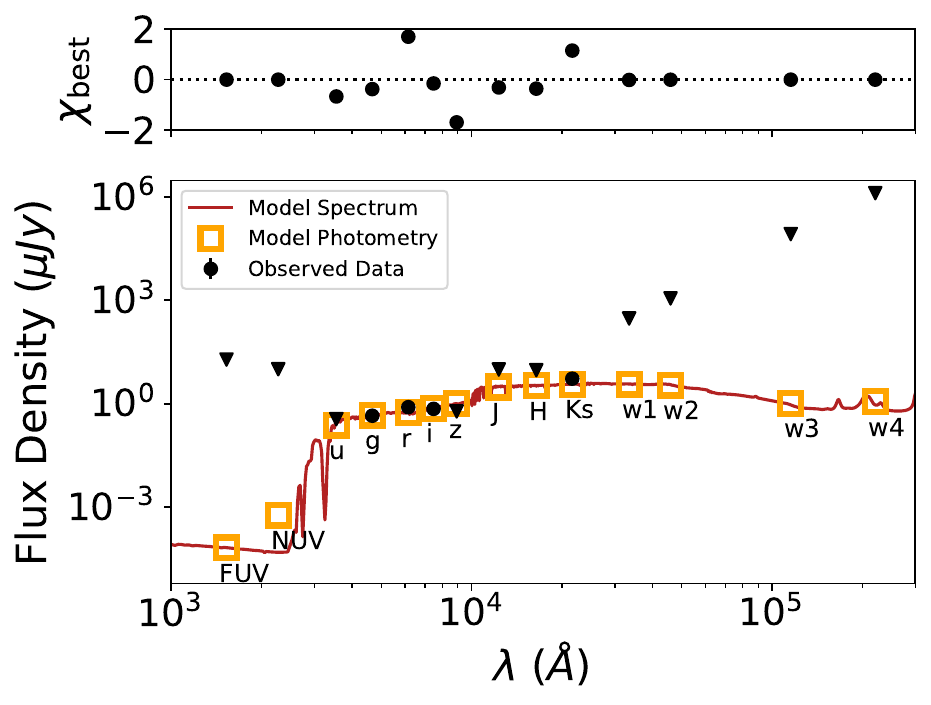}
\caption{SED of the host galaxy of EP240801a. The best-fit model SED is shown by the red line. The $\chi_{best}$ means standardized residual. The model photometry results are marked as squares with their corresponding bands below, and the observed data is indicated by black data points (black inverted triangles represent upper limits). All the data are corrected for Galactic extinction.
\label{host_spectrum}}
\end{figure}

\subsection{Host Galaxy}\label{sec:host}

A faint source located at (J2000)
R.A.~=~$23^{\rm hr}00^{\rm m}39.02^{\rm s}$, Dec.~=~$+32^\circ 35' 37.45''$
was detected in archival observations in the Legacy Surveys, which is $\sim 0.56''$ away from the position detected by TRT. According to the Legacy Surveys' result, the photometric redshift of this source is $0.99 \pm 0.35$ \citep{2021MNRAS.501.3309Z}. Given its proximity to the location of EP240801a and its redshift, it is probable that this faint source is the host galaxy of EP240801a. 
We calculated the chance alignment probability that one source as bright as, or brighter than, this galaxy within this offset is found in the GTC/HiPERCAM $r'$-band image. First, we determined the number density of sources brighter than, or as bright as, the galaxy in a region of 60"x60" centered around EP240801a. Then, assuming Poisson statistics, we calculated that the chance alignment for this galaxy is $P_{chance}$~=~0.002. Hence, it is likely that this is indeed the host galaxy of EP240801a.
Photometry of the host galaxy in the $g',\,r',\,i',\,z',\,J,\,Ks$ bands was obtained recently from GTC, NOT, SAI-25 and the 8.2~m Very Large Telescope (VLT) as listed in Table \ref{tab:optical_result}. We took the $u',\,H$ bands last detections by GTC as the upper limits of the host galaxy. We searched the catalogues and images in Galaxy Evolution Explorer All-Sky Survey (GALEX) and the Wide-field Infrared Survey Explorer final catalog (AllWISE) and calculated the 5$\sigma$ upper limits of these previous observations from the known sources in their catalogues or images to obtain the multiband photometry of the host galaxy. To investigate the stellar population properties of the host, we utilized \texttt{Prospector} \citep{prospector} to fit the host photometry; see Fig.~\ref{host_spectrum}. \texttt{Prospector} is a software for SED fitting to constrain the host-galaxy properties. The \textit{parametric sfh} template in \texttt{Flexible Stellar Population Synthesis} (\texttt{FSPS}, \citealt{Conroy_2009,Conroy_2010}) was used, and the redshift is fixed to 1.6734. The extinction of the host galaxy is assumed to be consistent with the Milky Way extinction law. The model photometry results in the $u',\,z',\,B,\,V,\,R,\,I,\,H$ bands were adopted to correct the contribution of the host galaxy for the optical data of EP240801a in those bands, and we adopted the Root Mean Squared Error (RMSE) as their error. The model photometric results for the $u'$ and $H$ bands may have an additional systematic error, given their wavelengths are located at the edges of the fitting data. Consequently, we presented all the observed data corrected for the host-galaxy contributions in Fig. \ref{totallc} while excluding the two $u'$- and $H$-bands data from the fits of afterglow models in Section \ref{sec:discussion}. The magnitudes of the host galaxy are shown in Table \ref{host_mag}.

\subsection{Closure Relations}
For the time period we are interested in (the first few days), the jet might enter the deceleration phase \citep{Sari1998}. The temporal decay index $\alpha = 0.95\,\pm\,0.03$ and the spectral index $\beta  =  0.69 \pm 0.02$, suggest that the frequencies of optical and X-ray emissions may lie between the minimum injection frequency $\nu_m$ and the synchrotron cooling frequency $\nu_{\rm c}$, namely $\nu_m \,  < \, \nu_{\rm optical} \,< \, \nu_{\rm X-ray} \, < \, \nu_c$. According to the closure relation \citep{CR04, CR13}, the electron energy distribution index $p$ is expected with $\beta  =  (p-1)/2$. As $p = 2.38\,\pm\,0.04$ derived from $\beta$, we have the temporal decay index $\alpha$ in a self-similar deceleration phase for $\nu_{\rm a} \,  < \, \nu_{\rm m} \,< \, \nu \, < \, \nu_{\rm c}$ and $p>2$ with:
\begin{equation}
F_\nu \propto
\begin{cases}
t^{-\frac{3(p-1)}{4}} & \text{ISM} \\
t^{-\frac{3p-1}{4}} & \text{Wind}
\end{cases}
\end{equation}
This corresponds to $\alpha_{\rm ISM} = 1.035\,\pm\,0.03$ and $\alpha_{\rm Wind} = 1.535\,\pm\,0.03$. The wind case is clearly excluded owing to the significant discrepancy with the observed temporal decay index $\alpha$.

\subsection{Theoretical Interpretation} \label{sec:theory}

The optical afterglow light curve (R band) exhibits two epochs, i.e., a shallow phase ($\alpha \sim 0.25$), followed by normal decay ($\alpha \sim 0.95$). Such behaviors imply complex jet emission components in EP240801a. To understand the afterglow data of EP240801a, we consider three different models, i.e., a two-component jet model, a reverse-forward shock model and a forward shock model with energy injection.

The closure relation analysis for EP240801a indicates an ISM environment (a constant external medium). We thus consider a relativistic shell with energy $E_{\rm 0}$, initial bulk Lorentz factor $\Gamma_0$, and opening angle $\theta_{\rm j}$, advancing in an ISM. A pair of shocks will be produced  \citep{Rees1992,Meszaros1997,Sari1998,Sari1999,Kobayashi1999,Zou2005,Lamb2019}: a forward shock (FS, propagating into the external medium) and a reverse shock (RS, propagating into the shell). There are four regions separated by the two shocks: Region 1, the ISM with density $n_1$; Region 2, the shocked medium; Region 3, the shocked shell material; and Region 4, the unshocked shell material with density $n_4$.

For the FS, the dynamics are described with four phases: the coasting phase (the Lorentz factor is nearly constant, $\Gamma\approx \Gamma_0$), the deceleration phase (the shell starts to decelerate when the mass of the medium swept by the FS is about $1/\Gamma_0$ of the rest mass in the ejecta), the post-jet-break phase (when the $1/\Gamma$ cone becomes larger than $\theta_{\rm j}$), and the Newtonian phase (when the rest-mass energy of the swept-up medium  becomes comparable to the energy of the ejecta). The overall evolution of the shell, covering the above four phases, is determined by \citep{Huang2000} \footnote{The equation (\ref{eq.dgamma}) is an approximate description of the blastwave dynamics. A rigorous treatment after correctly describing the internal energy, adiabatic loss and total energy should be \citep{Nava2013,Zhang2018book}
\begin{equation}
   \frac{d\Gamma}{dR} = -\frac{(\Gamma_{\rm eff}+1) (\Gamma-1)c^2 \frac{dm}{dR} + \Gamma_{\rm eff} \frac{dU_{\rm ad}}{dr} }{(M_{\rm ej} +m) c^2+ U \frac{d\Gamma_{\rm eff}}{d\Gamma} } \nonumber ,  
\end{equation}
where $\Gamma_{\rm eff}=(\hat{\gamma}\Gamma^2-\hat{\gamma}+1)/\Gamma$, $U$ is the internal energy in the comoving frame, $dU_{\rm ad}$ is the adiabatic loss. This equation can be reduced to equation (\ref{eq.dgamma}) by adopting $\hat{\gamma}=1$ (neglecting the pressure term), $dU_{\rm ad}/dR=0$ (neglecting adiabatic loss), $U=(\Gamma-1) mc^2$ and $\Gamma_{\rm eff}\simeq \hat{\gamma} \Gamma$ (for $\Gamma \gg 1$).}
\begin{equation}
\frac{dR}{dt} =\beta_{\rm j} c \Gamma (\Gamma+\sqrt{\Gamma^2 -1}),    
\end{equation}
\begin{equation}
\frac{dm}{dR} =2\pi R^2 (1-\cos\theta_{\rm j})n_1 m_{\rm p},   
\end{equation}
\begin{equation}\label{eq.dgamma}
\frac{d\Gamma}{dm} = -\frac{\Gamma^2 -1}{M_{\rm ej} + 2\Gamma m} , 
\end{equation}
where $R$ is the radius of the event in the burst frame, $t$ is the observer time, $m$ is the swept-up mass, $M_{\rm ej}=E_{\rm 0} (1-\cos\theta_{\rm j})/2(\Gamma_0-1)c^2$ is the ejecta mass, $m_{\rm p}$ is the proton mass, and $\beta_{\rm j}=\sqrt{1-\Gamma^{-2}}$. 

If there is energy injection from the GRB central engine, Eq. \ref{eq.dgamma} should be replaced by \citep{Geng2013}
\begin{equation}
\frac{d\Gamma}{dm} = -\frac{\Gamma^2 -1- \frac{1-\beta_{\rm j}}{\beta_{\rm j} c^3}L_{\rm inj} dR/dm}{M_{\rm ej} + 2\Gamma m} .   
\end{equation}
During the injection time, $t_{\rm start}<t<t_{\rm end}$, the injected luminosity is $L_{\rm inj}=L_{\rm inj}^0 (t/t_{\rm start})^{-q}$, where $L_{\rm inj}^0$ is the initial injection power, $q$ is the decay power-law index, and $t_{\rm start}$ and $t_{\rm end}$ are respectively the start and end time for energy injection.

We assume that a constant fraction $\epsilon_{\rm e}$ of the FS energy $e_2=4\Gamma^2 n_1 m_{\rm p} c^2$ is deposited into electrons, accelerating them to a power-law distribution $N(\gamma_{\rm e}) \propto \gamma_{\rm e}^{-p}$. This defines the minimum injected electron Lorentz factor, $\gamma_{\rm m}=\frac{p-2}{p-1} \epsilon_{\rm e} (\Gamma-1)\frac{m_{\rm p}}{m_{\rm e}}$, where $m_{\rm e}$ is the electron mass. A fraction $\epsilon_{\rm B}$ of the shock energy resides in the magnetic field with $B=(32\pi m_{\rm p} \epsilon_{\rm B} n_0)^{1/2} c$. The critical electron Lorentz factor $\gamma_{\rm c} = (6\pi m_{\rm e}c)/(\Gamma \sigma_{\rm T}B^2 t)$ is given by setting the electron's lifetime equal to the time $t$, and electrons with  $\gamma_{\rm e} >\gamma_{\rm c}$ will be significantly cooled due to synchrotron radiation.

Combining the radiative cooling and the continuous injection of new accelerated electrons, coupling with the synchrotron self-absorption effect, leads to a broken power-law spectrum. This spectrum is segmented into several sections based on three characteristic frequencies \citep{Sari1998,Chevalier_2000,Granot_2002,Wu_2003,Kobayashi_2003,Zou2005,Eerten_2009,CR13,Zhang2018book}: $\nu_{\rm m}$ (defined by $\gamma_{\rm m}$), $\nu_{\rm c}$ (defined by $\gamma_{\rm c}$), and $\nu_{\rm a}$ (characterized by synchrotron self-absorption).

Considering the contribution of the reverse shock, a ``rebrightening'' feature (a distinct RS peak and a FS peak) is expected in the optical light curve \citep{Zhang2003} for a thin-shell interacting with a constant-density ISM. Therefore, the optical afterglow behavior ($R$ band)  of EP240801a can also be explained with the thin-shell forward-reverse shock (FS-RS) model.

In the thin-shell case, the RS is Newtonian during the shock-crossing phase \citep{Kobayashi2000, Zhang2003}. The scalings before the RS crossing time $t_{\rm dec}$ are \citep{Kobayashi2000} 
\begin{eqnarray}
\gamma_3 \approx \Gamma_0,
\  n_3\approx 7 n_1 \Gamma_0^2(t/t_{\rm dec})^{-3} , \nonumber \\
\ e_3\approx 4 \Gamma_0^2 n_1 m_{\rm p} c^2,
\ N_{\rm e, 3} \approx N_0 (t/t_{\rm dec})^{3/2},   
\end{eqnarray}
where $N_0=M_{\rm ej}/m_{\rm p}$ is the total number of electrons in the ejecta. 

After the RS crosses the shell, the shell expands adiabatically in the shell's comoving frame, and the jet enters the deceleration phase. The dynamical behavior in Region 3 is expressed with the scalings \citep{Kobayashi2000, CR13,Zhang2018book}
\begin{equation}
\gamma_3 \propto t^{-2/5}, \ n_3 \propto t^{-6/7}, \ e_3 \propto t^{-8/7}, \ N_{\rm e,3} =N_0.    
\end{equation}

In the same way as in the FS, we also assume that electrons are accelerated at the RS front to a power-law distribution, a fraction of the RS energy $e_3$ is distributed into electrons, and a fraction to the magnetic field in Region 3. The spectrum is also segmented into a broken power law by $\nu_{\rm m}$, $\nu_{\rm c}$, and $\nu_{\rm a}$.

To explain the afterglow data of EP240801a with the three models (that is, a two-component jet model, an FS-RS model and an FS model with energy injection), we employ a numerical code \texttt{PyFRS}\footnote{\url{https://github.com/leiwh/PyFRS/}} to calculate the RS and FS emissions \citep{CR13,Zhang2018book, Zhu2023,Zhou2024,Fu2024}. We use the top-hat jet-type in the modeling. Markov chain Monte Carlo (MCMC) with \texttt{PyFRS} is adopted for multiband fitting to place constraints on model parameters. The MCMC fit is performed using the Python package \texttt{emcee} \citep{emcee2013}, which utilizes a group of parallel-tempered affine invariant walkers to explore the parameter space.

\begin{figure*}[ht!]
\gridline{\fig{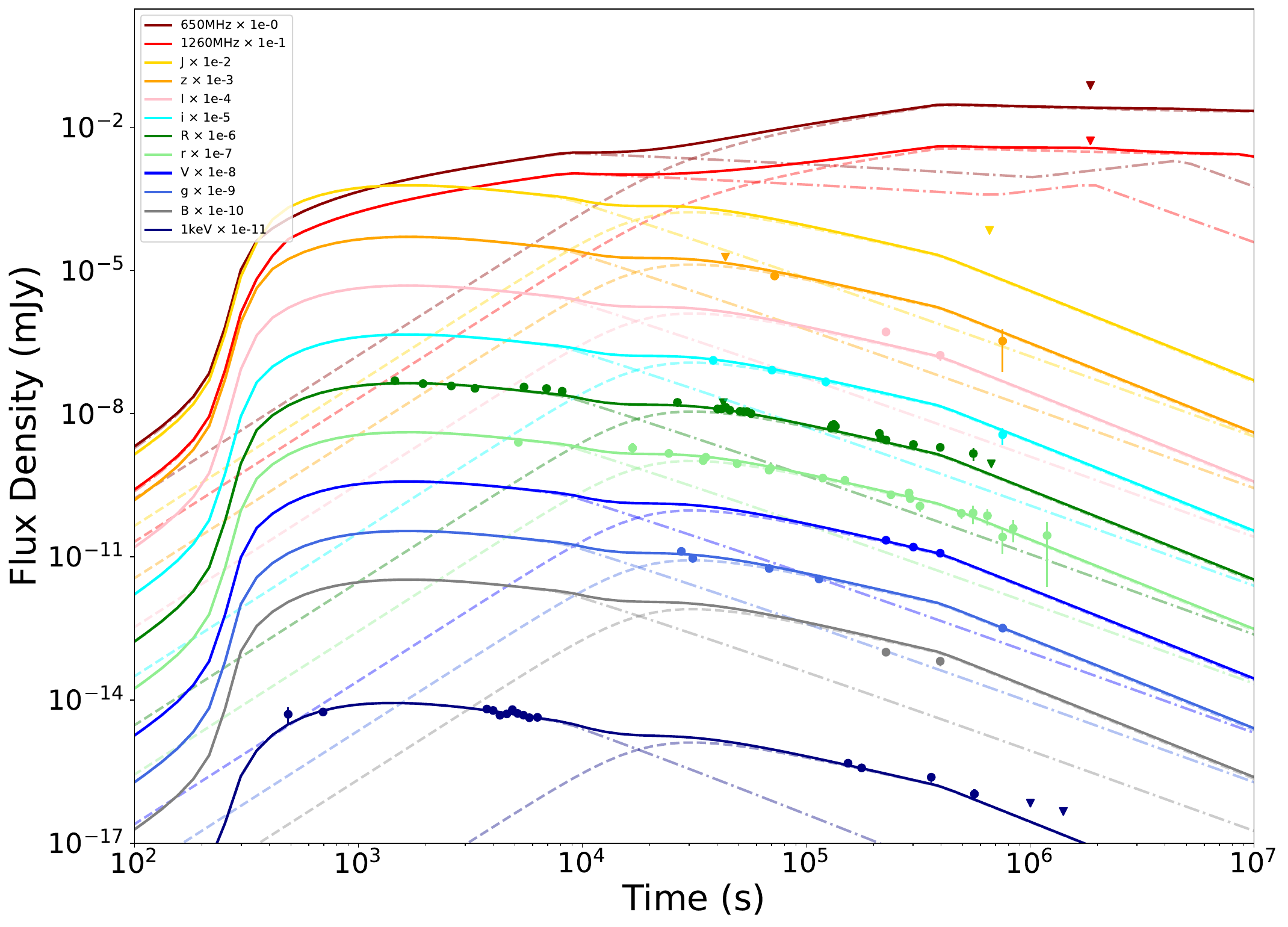}{0.7\textwidth}{(a)}}
\gridline{\fig{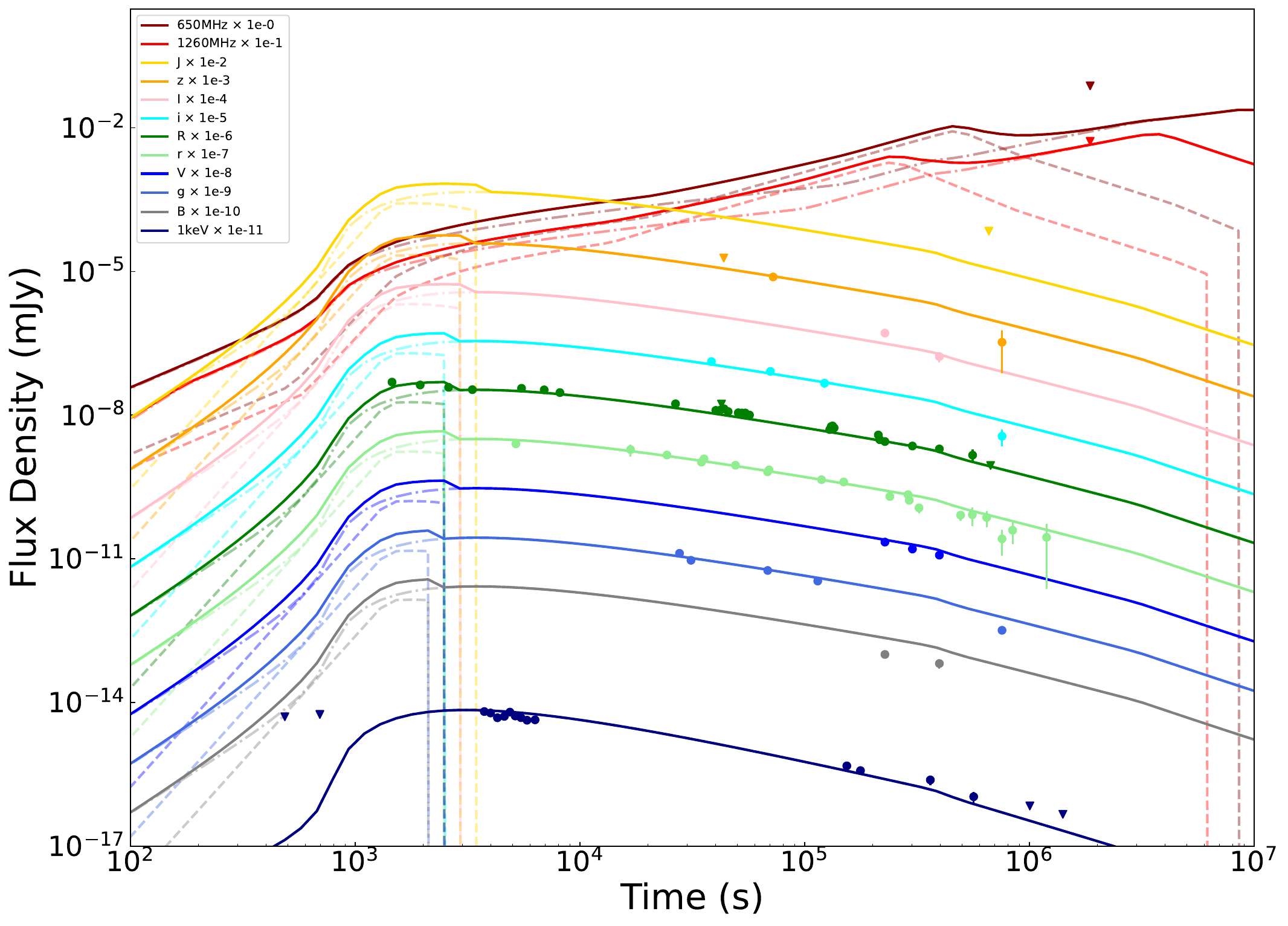}{0.5\textwidth}{(b)}
          \fig{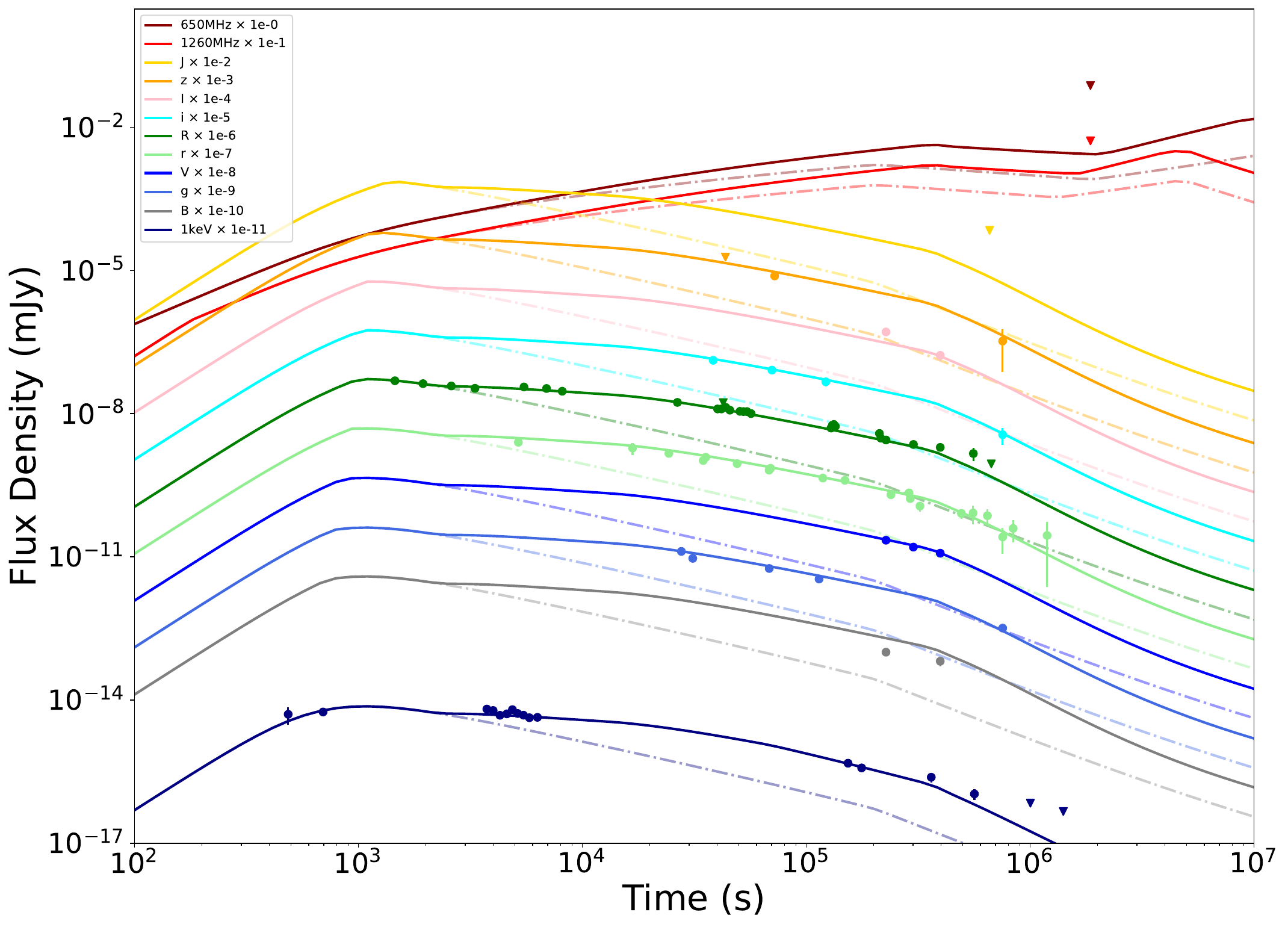}{0.5\textwidth}{(c)}
          } 
                   
\caption{Best fit results using \texttt{PyFRS} code with three models (solid lines): (a) two-component jet model, containing a narrow and faster jet (dash-dotted lines) as well as a broad, slower jet (dashed lines); (b) FS-RS model. The contributions from the RS and FS are plotted with dashed lines and dash-dotted lines, respectively; (c) FS with energy injection (solid lines). The forward shock without energy injection are represented by the dash-dotted line.\label{FitLC}}
\end{figure*}

\begin{center}
\begin{deluxetable*}{ccccccccc}[htbp!]
\tabletypesize{\scriptsize}
\tablecaption{Parameters of afterglow modeling
\label{tb.AGresults}}
\tablewidth{0pt}
\tablehead{
\colhead{Parameters} & \colhead{Unit} & \colhead{Prior Type} & \colhead{Prior} & 
\multicolumn{5}{c}{Results}\\
\cline{5-9}
\colhead{} & \colhead{} & \colhead{} & \colhead{} &  \multicolumn{2}{c}{Two-component Jet} & \multicolumn{2}{c}{FS+RS} & \colhead{Energy Injection}\\
\cline{5-9}
\colhead{} & \colhead{} & \colhead{} & \colhead{} &  \colhead{Narrow} & \colhead{Wide} &  \colhead{FS} & \colhead{RS} & \colhead{} 
}
\startdata
$n_{18}$ & cm$^{-3}$ & log-uniform& $[10^{-5}, 10^3] $& \multicolumn{2}{c}{$(2.24^{+6.27}_{-1.70})\times 10^{-3}$}  &  \multicolumn{2}{c}{$5.37_{-3.82}^{+15.05}$}  & $8.13^{+5.05}_{-4.24}$ \\
$\theta_{\rm obs}$ & degrees & uniform & $[0.01, 45]$ & \multicolumn{2}{c}{$1.50_{-0.35}^{+0.45}$} &   \multicolumn{2}{c}{$10.51_{-1.51}^{+2.01}$} & $8.01^{+5.15}_{-5.08}$ \\
$E_{\rm 0}$  & erg & log-uniform & $[10^{51}, 10^{56}]$ &  $(2.40^{+10.48}_{-1.89})\times10^{53}$ & $(8.51^{+23.11}_{-5.20}) \times 10^{52}$ & \multicolumn{2}{c}{$(1.58^{+0.28}_{-0.44}) \times 10^{53}$} & $(5.01^{+2.40}_{-1.12}) \times 10^{51}$ \\
$\Gamma_0$  & 1 & log-uniform & $[1, 2000]$ & $1258.93^{+439.32}_{-427.16}$  & $44.67^{+14.22}_{-10.00}$    & \multicolumn{2}{c}{$199.53_{-41.04}^{+40.36}$} & $40.74^{+6.04}_{-3.58} $ \\
$\theta_{\rm j}$ & degrees & uniform & $[0.01, 45]$ & $1.15^{+0.35}_{-0.27}$  & $5.25^{+1.49}_{-1.31}$    & \multicolumn{2}{c}{$9.03_{-1.30}^{+1.73}$} & $15.73_{-1.62}^{+1.30}$ \\
$p$ & uniform & 1 &  $[2.01, 4.0]$ &  $2.28_{-0.01}^{+0.01}$ &  $2.43_{-0.02}^{+0.02}$  &  $2.05^{+0.02}_{-0.02}$ & $2.47^{+0.37}_{-0.20}$  &  $2.35_{-0.02}^{+0.02}$  \\
$\epsilon_{\rm e}$ & 1 & log-uniform & $[10^{-4}, 0.5]$ & $(6.31^{+10.29}_{-4.72}) \times 10^{-2}$  & $(3.72^{+0.86}_{-0.96}) \times 10^{-1}$ & $(4.37^{+1.80}_{-0.90}) \times 10^{-2}$ & $(3.39^{+0.50}_{-0.82}) \times 10^{-1}$  & $(3.89^{+0.68}_{-1.01}) \times 10^{-1}$ \\
$\epsilon_{\rm B}$ & 1 & log-uniform & $[10^{-4}, 0.5]$ & $(2.57^{+25.61}_{-2.47}) \times 10^{-3}$  & $(1.91^{+8.09}_{-1.66}) \times 10^{-3}$ & $(7.41^{+9.18}_{-4.25}) \times 10^{-4}$ & $(9.33^{+135.21}_{-8.42}) \times 10^{-3}$   & $(1.38^{+0.66}_{-0.28}) \times 10^{-4}$ \\
$L_0$ & $\rm erg\,s^{-1}$ & log-uniform & $[10^{45}, 10^{54}]$ & --  & --  & -- & --  & $(1.38_{-0.47}^{+0.81}) \times 10^{48}$ \\
$t_0$ & s & log-uniform & $[10^{3}, 10^{4}]$ & -- & --  & --  & --  & $977.24_{-218.66}^{+144.78}$ \\
$t_{\rm e}$ & s & log-uniform & $[10^{3}, 10^{7}]$ & -- & --  & -- & --  & $4897.79_{-1095.89}^{+1709.15}$ \\
$q$ & 1 & uniform & $[-2,2]$ & -- & --  & --  & --   & $0.11_{-0.35}^{+0.27}$ \\
\enddata
\end{deluxetable*}
\end{center}

The early steep decay phase at several 100 s, as observed by EP/FXT, is not included in the afterglow fit, as this data is likely dominated by the prompt emission. However, the two data points at the end of the steep decline deviate from the single powerlaw fit to this data ($\alpha \sim 5.82\pm0.32$), and are thus considered in the afterglow fit. The two $u'$- and $H$- bands data are excluded from the fit, as their potential systematic errors whether in the differential photometry or the host-galaxy model photometric results mentioned above.

In our MCMC fitting with \texttt{PyFRS}, we set the walkers as tenfold the number of free parameters (i.e. 140 for the two-component jet model) running 30,000 steps and discarded the first 15,000 steps as burn-in to explore the parameter space. The best-fitting light curve and the posterior probability distributions with 1$\sigma$ uncertainties of the parameters are illustrated in Figs. \ref{FitLC} and listed in Table \ref{tb.AGresults}, respectively. As we can see, all these three models can generally reproduce the multiband afterglow observations. We will discuss the fit results in the next section.

\section{Discussion}\label{sec:discussion}
\subsection{Two-Component Jet Model Fit}\label{sec:2FS}
A two-component jet model can produce distinct behavior in the afterglow light curve. In the fit, the narrow component and the wide component are denoted with the subscripts ``n'' and ``w'', respectively. 

Fourteen free parameters are included in the fitting: the isotropic kinetic energies $E_{\rm 0,n}$ and $E_{\rm 0,w}$, the initial Lorentz factors $\Gamma_{\rm 0, n}$ and $\Gamma_{\rm 0, w}$, the jet opening angles $\theta_{\rm j, n}$ and $\theta_{\rm j, w}$, the viewing angle $\theta_{\rm obs}$, the number density of the ISM $n_{1}$ ($n_{18}$ in the \texttt{PyFRS} code, defined as the density at $R=10^{18}$ cm), the electron distribution power-law indices $p_{\rm n}$ and $p_{\rm w}$, the energy fractions in electrons $\epsilon_{\rm e, n}$ and $\epsilon_{\rm e, w}$, and the energy fractions in the magnetic field $\epsilon_{\rm B, n}$ and $\epsilon_{\rm B, w}$.

\begin{figure*}[ht!]
\gridline{\fig{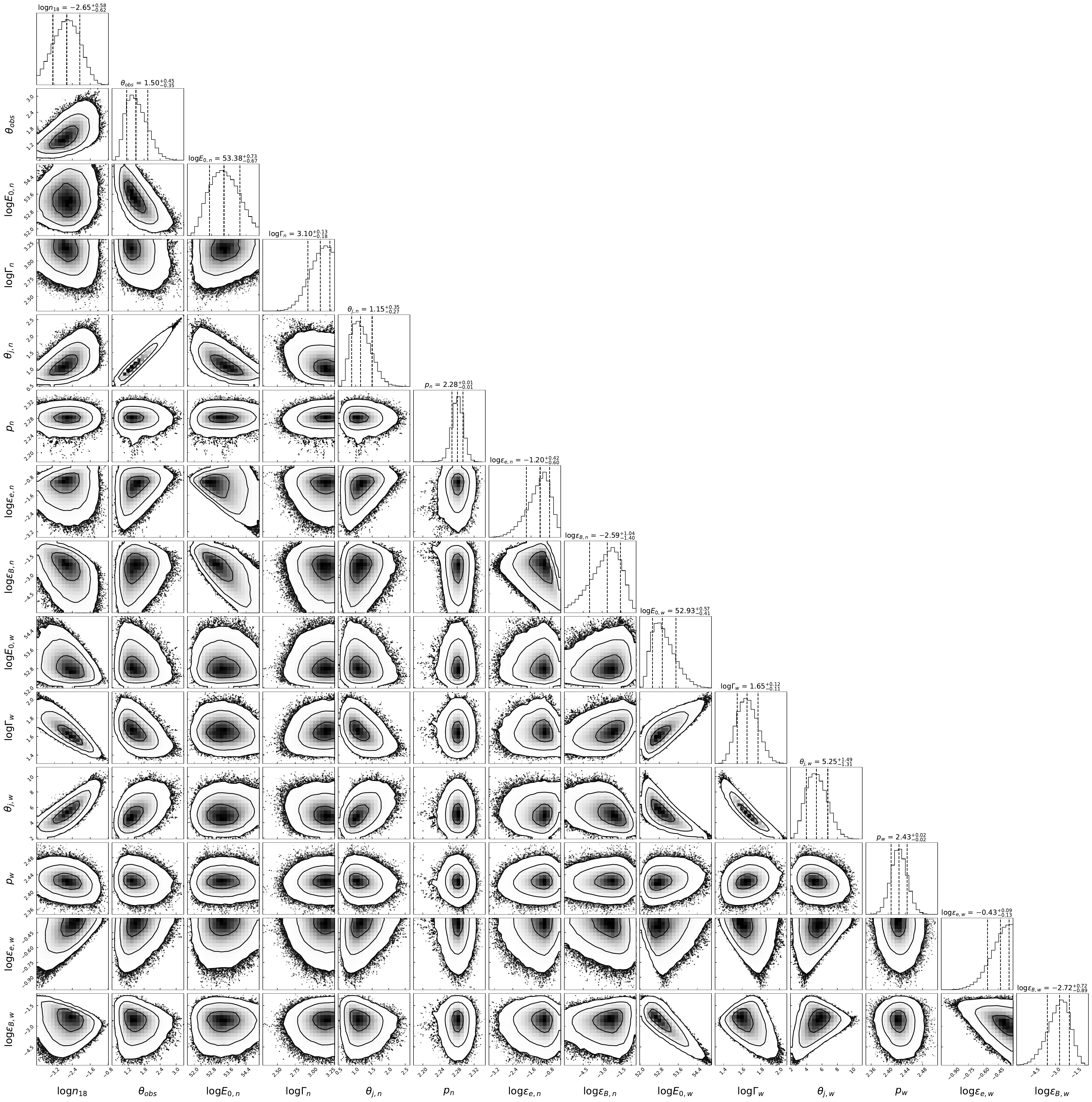}{0.8\textwidth}{}
          %\fig{fig/FRS_k0_241212_1_corner.pdf}{0.45\textwidth}{(b)}
          }
\caption{Posterior probability distributions of the afterglow parameters were obtained using the two-component jet modeling of EP240801a. The median values with the $1\sigma$ regions are shown in the 1D probability distribution.\label{2FScorner}}
\end{figure*}

The best-fitting result and the posterior probability distributions of the parameters are presented in Table \ref{tb.AGresults} and illustrated in Fig.~\ref{FitLC}(a) and \ref{2FScorner}, respectively.

We find a narrow-component jet with $E_{\rm 0, n} \approx 2.4\times 10^{53}$ erg, $\Gamma_{\rm 0,n}\approx 1259$, $\theta_{\rm j,n} \approx 1.15^\circ$, $p_{\rm n} \approx 2.28$, $\epsilon_{\rm e,n} \approx 6.31\times 10^{-2}$ and $\epsilon_{\rm B,n} \approx 2.57\times 10^{-3}$, and a wide-component jet with $E_{\rm 0, w} \approx 8.5\times 10^{52}$ erg, $\Gamma_{\rm 0,w}\approx 45$, $\theta_{\rm j,w} \approx 5.25^\circ$, $p_{\rm w} \approx 2.43$, $\epsilon_{\rm e,w} \approx 3.72\times 10^{-1}$ and $\epsilon_{\rm B,w} \approx 1.91\times 10^{-3}$. We find the ISM number density of $n_{1} \approx 2.24 \times 10^{-3}\, \rm cm^{-3}$, and a viewing angle of $\theta_{\rm obs} \approx 1.5^\circ$. The early optical and X-ray afterglows are dominated by the narrow jet, while the late afterglows are dominated by the wide jet. 

The narrow jet is strong, with $E_{\rm 0, n} \approx 2.4\times 10^{53}$ erg and $\Gamma_{\rm 0,n}\approx 1259$. Typically, a normal GRB is expected. However, our line-of-sight (LoS) deviates slightly from the narrow jet beam \footnote{The beam angle is given by ${\rm max}(\theta_{\rm j}, 1/\Gamma)$.The narrow-beam jet is viewed off-axis if the observing angle $\theta_{\rm obs}>{\rm max}(\theta_{\rm j,n}, 1/\Gamma_{\rm 0, n})$.}, the prompt emission will be significantly suppressed due to the strong Doppler beaming. To test this point, we consider an off-axis observer at $\psi$ relative to the edge of the jet, the observed flux density could be expressed as \citep{off-axis_correct_one,Lei2016,Ioka2018,Beniamini2023,Lamb2019b}
\begin{eqnarray}
    F_{\nu}(\psi,t)\!\sim\! F_{\nu /a_{\rm off}}(0,a_{\rm off} t)\!\left\{ \begin{array}{ll}  \!a_{\rm off}^2 &; \theta_{\rm j}\!<\!\theta_{\rm obs}\! \leq \!2\theta_{\rm j}\ ,\\ \!(\Gamma\theta_{\rm j})^2a_{\rm off}^3 &; \theta_{\rm obs}\!>\!2\theta_{\rm j}\ ,\\ \!a_{\rm off}^3 &; \theta_{\rm j}\!\leq\!\Gamma^{-1}\ ,\end{array} \right.
\label{eq:Fvoff}
\end{eqnarray}
% \begin{equation}
% F_{\nu}(\psi,t) = a_{\rm off}^3 F_{\nu /a_{\rm off}}(0,a_{\rm off} t),
% \label{eq:Fvoff}
% \end{equation}
where $a_{\rm off}={\mathcal D}_{\rm off}/{\mathcal D}_{\rm on}=
(1-\beta_{\rm j})/(1-\beta_{\rm j} \cos \psi)$ is the ratio of the on-beam to the off-beam Doppler factor.

With the viewing angle $\theta_{\rm obs} \approx 1.5^\circ$,  the Lorentz factor $\Gamma_{\rm 0,n} \approx 1259$ and opening angle $\theta_{\rm j,n} \approx 1.15^\circ$ of the narrow jet, we correct the gamma-ray energy $E_{\gamma,\rm{iso}}$ and $E_{\rm peak}$ to the on-axis situation by using \citep{off-axis_correct_one,off-axis_correct_two,Ioka2018} $E_{\gamma,\rm{iso}}^{\rm off} /E_{\gamma,\rm{iso}}^{\rm on} \approx a_{\rm off}^2$ and $ E_{\rm{peak}}^{\rm off}/E_{\rm{peak}}^{\rm on} \approx a_{\rm off}$. One can derive the on-axis gamma-ray energy $E_{\gamma,\rm{iso}}^{\rm on} \sim 2.0 \times 10^{55}\,\rm {erg}$, and $E_{\rm{peak}}^{\rm on} \sim 895\,\rm{keV}$, the peak energy in the rest frame is 2393 keV with $z=1.6734$. Note that for ``BOAT'' GRB 221009A, the investigations also reveal a physical picture involving two jet components, i.e., a narrow pencil-beam jet and a broder jet wing \citep{Zhang2024,Rhodes2024}. The above on-axis $E_{\gamma,\rm{iso}}^{\rm on}$, and $E_{\rm{peak}}^{\rm on}$ of EP240801a are comparable to those of GRB 221009A ($(1.5 \pm 0.2) \times 10^{55} \rm erg$ and $1435 \pm 105$ keV in the rest frame, from \citealt{GRB221009A_GECAM}). Therefore, EP240801a could be similar to GRB 221009A, but viewed slightly off-axis. We are observing the off-axis narrow beam emission, making EP240801a an X-ray flash, if the prompt emission is mianly produced by the narrow jet.

Besides the narrow beam emission, the wide component contains small values of the kinetic energy ($E_{\rm 0, w} \approx 8.5\times 10^{52}$ erg) and the initial Lorentz factor ($\Gamma_{\rm 0,w}\approx 45$). It is also possible that the bright core is missed by the observer, and the prompt emission of EP240801a is mainly produced by the weak wide jet.

\subsection{Forward-reverse Shock Model Fit}\label{sec:FSRS}

In a thin shell FS-RS model, a ``rebrightening'' feature (a distinct RS peak and a FS peak) is expected in the optical light curve \citep{Zhang2003}. We, therefore, also consider this model to fit the afterglow of EP240801a.

The FS and RS components are denoted with the subscripts ``f'' and ``r'', respectively. Eleven free parameters are included in the fitting: the isotropic kinetic energy $E_{\rm 0}$, the initial Lorentz factor $\Gamma_{\rm 0}$, the jet opening angle $\theta_{\rm j}$, the viewing angle $\theta_{\rm obs}$, the number density of the ISM $n_{1}$, and the microphyscis parameters for the FS ($p_{\rm f}$, $\epsilon_{\rm e,f}$, $\epsilon_{\rm B,f}$) and RS ($p_{\rm r}$, $\epsilon_{\rm e,r}$, $\epsilon_{\rm B,r}$).

The two EP/FXT data points at the end of the steep decline can not be reproduced with this model, we just set them as upper limits in the afterglow fit in this case. The best-fitting result are presented in Table \ref{tb.AGresults} and illustrated in Fig.~\ref{FitLC}(b). 

We find an ISM number density of $n_{1} \approx 5.37  \rm cm^{-3}$, a viewing angle of $\theta_{\rm obs} \approx 10.5^\circ$, an isotropic kinetic energy $E_{\rm 0} \approx 1.58\times 10^{53}$ erg,  an initial Lorentz factor $\Gamma_{\rm 0}\approx 200$, and an opening angle $\theta_{\rm j} \approx 9^\circ$. The values of the FS (RS) microphysics parameters are  $p_{\rm f} \approx 2.05$ ($p_{\rm r} \approx 2.47$), $\epsilon_{\rm e,f} \approx 4.4\times 10^{-2}$ ($\epsilon_{\rm e,r} \approx 3.4\times 10^{-1}$), and $\epsilon_{\rm B,f} \approx 7.4\times 10^{-4}$ ($\epsilon_{\rm B,r} \approx 9.3\times 10^{-3}$).

The FS-RS model fit result also implies that EP240801a is observed off-axis ($\theta_{\rm obs} \approx 10.5^\circ$ vs. $\theta_{\rm j} \approx 9^\circ$). In the same way as done in Section \ref{sec:2FS}, we correct the  the gamma-ray energy $E_{\gamma,\rm{iso}}$ and also $E_{\rm peak}$ to the on-axis situation. One can derive the on-axis gamma-ray energy $E_{\gamma,\rm{iso}}^{\rm on} \sim 4.3 \times 10^{54}\,\rm {erg}$, and $E_{\rm{peak}}^{\rm on} \sim 412\,\rm{keV}$, appeared as a bright Type II GRB.

A magnetically dominated jet does not provide a reverse shock \citep{Zhang2005}. If the eary phase of R-band light curve is shaped by the FS-RS emission, then the RS should not be strongly magnetized. Based on the fit with the FS-RS model, we infer that the magnetic field strength ratio in the RS and FS is $R_{\rm B}\equiv B_{\rm r}/B_{\rm f} \approx 3.5$. Such a mild $R_{\rm B}$ is consistent with the forward-reverse shock interpretation.

\subsection{Energy Injection Model Fit} \label{sec:EngInj}
The X-ray and optical shallow decay can also be interpreted in the FS model with energy injection due to late-time central engine activity. The best-fitting results are presented in Table \ref{tb.AGresults} and illustrated in Fig.~\ref{FitLC}(c). 

We find a jet with $E_{\rm 0} \approx 5.01\times 10^{51}$ erg, $\Gamma_{\rm 0}\approx 41$, $\theta_{\rm j} \approx 15.73^\circ$, $p_{\rm} \approx 2.35$, $\epsilon_{\rm e} \approx 3.89\times 10^{-1}$, and $\epsilon_{\rm B} \approx 1.38\times 10^{-4}$. We find an ISM number density of $n_{1} \approx 8.13$, and viewing angle of $\theta_{\rm obs} \approx 8.01^\circ$, energy injection with luminosity $L_{0} \approx 1.38\times 10^{48}\, \rm erg\, \ s^{-1}$, power-law index $q\approx 0.11$, start time $t_{\rm 0} \approx 977$ s, and end time $t_{\rm e}\approx 4898$ s. 

We consider two popular models for the energy injection: spin-down of a magnetar and fall-back accretion onto a stellar mass BH. The characteristic spin-down luminosity $L_0$ of a magnetar is \citep{Zhang2001}
\begin{equation}
L_0 = 1.0\times 10^{49}\, \left(B_{p,15}^2 P_{0,-3}^{-4}R_6^6\right)\, \rm{erg\, s^{-1}},
\end{equation}
where $B_{\rm p,15}=B_{\rm p}/(10^{15}\, \rm G$, $P_{0,-3}=P_0/(10^{-3}\, \rm ms)$, and $R_6=R/(10^6\, \rm cm)$. $B_{\rm p}$, $P_0$, and $R$ are the magnetic field, spin period, and radius of the magnetar. 

For the BH central engine model, late energy injection is dominated by the BZ mechanism \citep{Liu2015, Lei2017}. The BZ power can be rewritten as a function of mass accretion rate as \citep{Lei2013}
\begin{equation}
L_{\rm BZ}=9.3 \times 10^{48} \frac{a_\bullet^2 \dot{m}_{-5}  F(a_\bullet)}{(1+\sqrt{1-a_\bullet^2})^2} \ {\rm erg \ s^{-1}} ,
\label{eq:EB}
\end{equation}
where $\dot{m}_{-5} = \dot{M}/(10^{-5} M_\sun\ \rm s^{-1})$ is the dimensionless accretion rate, $a_\bullet=J_\bullet c/(GM_\bullet^2)$ is the spin parameter of the BH, $F(a_{\bullet})=[(1+q_a^2)/q^2][(q_a+1/q_a) \arctan q_a-1]$, and $q_a= a_{\bullet} /(1+\sqrt{1-a^2_{\bullet}})$.

As we can see, both central engine models can give rise to the energy injection required for typical values of the parameters of the objects involved. The values of jet parameters are quite close to those of the wide jet of two-component jet model, i.e., kinetic energy $E_{\rm 0} \sim 10^{52}$ erg and Lorentz factor $\Gamma_0 <50$, making it possible to produce a X-ray flash like EP240801a.

\section{Summary}\label{sec:summary}
In this paper, we present comprehensive multiband observations with the facilities mentioned in Appendix \ref{sec:photometic table} of EP240801a, an extremely soft GRB detected by EP. The physical origin of such a burst is still unclear. Our rich dataset, especially the early X-ray and optical observations of EP240801a, enable us to explore its nature.

Our conclusions are summarized as follows: 

(1) We identified the redshift of EP240801a as $z = 1.6734\,\pm\,0.0002$ through the significant Mg~II doublet and Fe~II absorption lines that appear in the GTC and Keck spectra, assuming the transient occurred in the host galaxy at that redshift. 

(2) We performed a joint fit for the prompt emission phase with EP/WXT and Fermi/GBM data, deriving $E_{\gamma,\rm{iso}} = 5.57^{+0.54}_{-0.50}\times 10^{51}\,\rm{erg}$, $E_{\rm{peak}} = 14.90^{+7.08}_{-4.71}\,\rm{keV}$, and the ﬂuence ratio $\rm{S}(25-50\,\rm{keV})/\rm{S}(50-100\,\rm{keV})  =  1.67^{+0.74}_{-0.46}$ for EP240801a. These values suggest that EP240801a is an X-ray flash.

(3) We fit the available host galaxy photometry with \texttt{Prospector}, and use the best-fit model photometry to subtract the contribution of the host galaxy from the observed data.

(4) The $R$-band light curve shows a shallow phase followed by a normal decay phase, suggesting that multiple components are involved. Three models: a two-component jet model, an FS-RS model and a jet model with energy injection, are employed to elucidate the multiband afterglow data.

(i) The fit with the two-component jet model suggests an off-axis narrow jet and a weak wide jet. We find that EP240801a would be similar to the ``BOAT''  GRB 221009A if corrected to an on-axis view. Therefore, the X-ray flash EP240801a can be interpreted as the off-axis narrow beam emisision or the wide jet emission.

(ii) The FS-RS modeling also suggests an off-axis jet. The fit results for kinetic energy ($E_0 \sim 1.6\times 10^{53}$ erg), and initial Lorentz factor ($\Gamma \sim 200$) are consistent with a typical GRB. The modeling indicates that the RS/FS magnetic field strength ratio is $R_{\rm B} \equiv B_{\rm r}/B_{\rm f} \approx 3.5$. Such a mild $R_{\rm B}$ is also consistent with the forward-reverse shock interpretation.

(iii) The energy injection model fit also involves a weak jet as the wide jet in the two-component jet model. Both the magnetar and the BH central engine models can explain the energy injection required for typical values of parameters. Therefore, EP240801a can also be interpreted as an intrinsically weak GRB.

Future EP detections might discover a number of X-ray flashes like EP240801a, which in turn would help to comprehend the physics behind such events. Especially, the very early X-ray and optical follow-ups, would be crucial to distinguish between the three origin models for EP240801a-like events, i.e., from a two-component jet similar to GRB 221009A whose bright narrow core is missed by the observer, from an off-axis normal GRB, or from a faint GRB (e.g., due to inefficient jet breakout from the progenitor star). These studies could be helpful in answering the question whether the narrow-beam GRBs like 221009A are common. 

\section*{Acknowledgments}
The data presented herein were obtained in part with ALFOSC, which is provided by the Instituto de Astrofisica de Andalucia (IAA) under a joint agreement with the University of Copenhagen and NOT. 
The optical spectrum was obtained at the W. M. Keck
Observatory, which is operated as a scientific partnership among the
California Institute of Technology, the University of California, and
NASA; the observatory was made possible by the generous financial
support of the W. M. Keck Foundation.
KAIT and its ongoing operation were made possible by donations from Sun Microsystems, Inc., the Hewlett-Packard Company, AutoScope Corporation, Lick Observatory, the U.S. NSF, the University of California, the Sylvia \& Jim Katzman Foundation, and the TABASGO Foundation. Research at Lick Observatory is partially supported by a generous gift from Google. 

D.X. acknowledges the science research grants from the China Manned Space Project with No. CMS-CSST-2021-A13 and CMS-CSST-2021-B11. This work is supported by the National Natural Science Foundation of China under grant 12473012.
A.V.F.'s research group at UC Berkeley acknowledges financial assistance from  the Christopher R. Redlich Fund, Gary and Cynthia Bengier, Clark and Sharon Winslow, Alan Eustace (W.Z. is a Bengier-Winslow-Eustace Specialist in Astronomy), William Draper, Timothy and Melissa Draper, 
Briggs and Kathleen Wood, Sanford Robertson (T.G.B. is a Draper-Wood-Robertson Specialist in Astronomy), and numerous other donors.   
T.-W.C. \& A.A. acknowledge the Yushan Young Fellow Program by the Ministry of Education, Taiwan for the financial support (MOE-111-YSFMS-0008-001-P1).
S.Y. acknowledges the funding from the National Natural Science Foundation of China under Grant No. 12303046 and
the Henan Province High-Level Talent International Training Program.
OAB and ASP thank the Ministry of Higher Education, Science and innovation of the Republic of Uzbekistan for financial support, project no. IL-5421101855.
FEB acknowledges support from ANID-Chile grants: BASAL CATA FB210003, FONDECYT Regular 1241005, and Millennium Science Initiative, AIM23-0001.
JQV acknowledges support from the European Union’s Horizon 2020 research and innovation programme Grant agreement No. 101095973 and the IAU-Gruber foundation fellowship.
MER acknowledges support from the European Union’s Horizon 2020 research and innovation programme Grant agreement No. 101095973.
AJCT acknowledges support from the Spanish Ministry projects PID2020-118491GB-I00 and PID2023-151905OB-I00 and Junta de Andaluc\'ia grant P20\_010168.
MAPT and DMS acknowledge support by the Spanish Ministry of Science via the Plan de Generacion de conocimiento PID2021-124879NB-I00. DMS also acknowledges support via a Ramon y Cajal Fellowship RYC2023-044941.

\facility{EP (WXT and FXT), Fermi (GBM), Keck (LRIS), KAIT, BOOTES-7, LCO, TRT, SLT, GMG, Jinshan-100C, ZTSh, AS-32, NOT (ALFOSC), BTA, AZT-20, AZT-22, SAI-25, GTC (OSIRIS+, HiPERCAM and EMIR), VLT (HAWK).}

\bibliography{main}{}
\bibliographystyle{aasjournal}

\appendix
\restartappendixnumbering

\section{The Photometric Results}\label{sec:photometic table}
\defcitealias{GCN37002}{GCN 37002}
\defcitealias{GCN37007}{GCN 37007}
\defcitealias{GCN37008}{GCN 37008}
\defcitealias{GCN37013}{GCN 37013}
\defcitealias{GCN37014}{GCN 37014}
\defcitealias{GCN37015}{GCN 37015}
\defcitealias{GCN37024}{GCN 37024}
\defcitealias{GCN37468}{GCN 37468}

\begin{ThreePartTable}
\begin{TableNotes}
\item[] $\Delta T$ is the exposure median time after the $T_0$. Magnitudes in the AB system \citep{ABsystem} are not corrected for Galactic extinction, which is $E(B - V) = 0.093$ \citep{SF2011}.
\item[*] We calibrated the magnitudes of the KAIT $Clear$-band with the $R$-band reference star magnitudes, as its effective wavelength is roughly $R$. For radio observations, the flux density unit is $\mu Jy$. The errors are statistical only.
\item[\dag] For these observations, they are mostly attributed by the host-galaxy flux, therefore we take them as host-galaxy magnitudes.
\end{TableNotes}
\setlength{\tabcolsep}{3pt}
\begin{center}
\begin{longtable}{ccccc}
    \caption{The Photometric Results of Our Observations
	Combined with Collected GCN Results}
	\label{tab:optical_result} \\
	\hline
	$\Delta T$ & Band& Magnitude\tnote{*} & Telescope & Reference \\ 
	(day) &  & (AB)& Inst. &  \\
	\hline
	\endfirsthead
	\multicolumn{5}{c}{{\textbf{\tablename\ \thetable{}.} Continued}} \\
    \hline
    $\Delta T$ & Band& Magnitude\tnote{*} & Telescope & Reference \\ 
    (day) &  & (AB)& Inst. &  \\
    \hline
    \endhead
    \hline
    \insertTableNotes  % 插入脚注
\endlastfoot
0.0169&  Clear&  19.89  $\pm$  0.07&  KAIT&  This work  \\
0.0225&  Clear&  20.04  $\pm$  0.06&  KAIT&  This work  \\
0.0301&  Clear&  20.16  $\pm$  0.06&  KAIT&  This work  \\
0.0331&  Clear&  $>$ 19.80 & BOOTES-7 & This work \\
0.0384&  Clear&  20.28  $\pm$  0.09&  KAIT&  This work  \\
0.0600&  r&  20.61  $\pm$  0.08&  LCO&  This work  \\
0.0635&  Clear&  20.21  $\pm$  0.08&  KAIT&  This work  \\
0.0801&  Clear&  20.29  $\pm$  0.09&  KAIT&  This work  \\
0.0941&  R&  20.43  $\pm$  0.17&  TRT&  This work  \\
0.1940&  r&  20.90  $\pm$  0.30&  SLT&  \citetalias{GCN37002}  \\
0.2818&  r&  21.17  $\pm$  0.11&  SLT&  This work  \\
0.3077&  R&  21.00  $\pm$  0.12&  GMG&  This work  \\
0.3204&  g&  21.43  $\pm$  0.20&  LCO&  \citetalias{GCN37007}  \\
0.3604&  g&  21.78  $\pm$  0.07&  ALT/100C&  This work  \\
0.4012&  r&  21.52  $\pm$  0.07&  ALT/100C&  This work  \\
0.4120&  r&  21.36  $\pm$  0.15&  SLT&  This work  \\
0.4445&  i&  21.25  $\pm$  0.09&  ALT/100C&  This work  \\
0.4649&  R&  21.33$\pm$0.04&  AZT-22&  This work  \\
0.4839&  R&  21.33  $\pm$  0.03&  ZTSh&  This work  \\
0.4928&  R&  $>$ 21.00&  AS-32&  This work  \\
0.5044&  z&  $>$ 20.70&  ALT/100C&  This work  \\
0.5058&  R&  21.25  $\pm$  0.02&  ZTSh&  This work  \\
0.5276&  R&  21.39  $\pm$  0.02&  ZTSh&  This work  \\
0.5687&  r&  21.70  $\pm$  0.03&  NOT&  This work \\
0.5852&  R&  21.45  $\pm$  0.02& ZTSh&  This work  \\
0.6071&  R&  21.46  $\pm$  0.02&  ZTSh&  This work  \\
0.6289&  R&  21.46  $\pm$  0.02&  ZTSh&  This work  \\
0.6563&  R&  21.56  $\pm$  0.02&  ZTSh&  This work  \\
0.7898&  r&  21.98  $\pm$  0.05&  GTC&  This work \\
0.7911&  g&  22.28  $\pm$  0.04&  NOT&  This work  \\
0.8019&  r&  21.89  $\pm$  0.04&  NOT&  This work  \\
0.8135&  i&  21.73  $\pm$  0.04&  NOT&  This work  \\
0.8363&  z&  21.69  $\pm$  0.09&  NOT&  This work  \\
1.3209&  g&  22.78  $\pm$  0.13&  ALT/100C&  This work  \\
1.3730&  r&  22.34  $\pm$  0.10&  ALT/100C&  This work  \\
1.4155&  i&  22.28  $\pm$  0.18&  ALT/100C&  This work  \\
1.4935&  R&  22.26  $\pm$  0.08&  ZTSh&  This work  \\
1.5155&  R&  22.12  $\pm$  0.07&  ZTSh&  This work  \\
1.5374&  R&  22.09  $\pm$  0.07&  ZTSh&  This work  \\
1.5620&  R&  22.18  $\pm$  0.08&  ZTSh&  This work  \\
1.7225&  r&  22.44  $\pm$  0.04&  NOT&  This work  \\
2.4565&  R&  22.51  $\pm$  0.12&  ZTSh&  This work  \\
2.4937&  R&  22.72  $\pm$  0.1&  AZT-22&  This work  \\
2.6284&  B&  23.90  $\pm$  0.08&  BTA&  This work  \\
2.6287&  I&  22.15  $\pm$  0.07&  BTA&  This work  \\
2.6287&  R&  22.80  $\pm$  0.05&   BTA&  This work  \\
2.6290&  V&  23.08  $\pm$  0.06&  BTA&  This work  \\
2.7656&  r&  23.03  $\pm$  0.07&  NOT&  This work  \\
3.3305&  r&  22.96  $\pm$  0.16&  ALT/100C&  This work  \\
3.3733&  r&  23.17  $\pm$  0.11&  AZT-20&  This work  \\
3.4843&  V&  23.36  $\pm$  0.07&  BTA&  This work  \\
3.4852&  R&  22.99  $\pm$  0.04&  BTA& This work  \\
3.7278&  r&  23.42  $\pm$  0.13&  NOT&  This work  \\
4.5923&  R&  23.11  $\pm$  0.05&  BTA&  This work  \\
4.5924&  I&  23.09  $\pm$  0.15&  BTA&  This work  \\
4.5925&  B&  24.21  $\pm$  0.10&   BTA&  This work  \\
4.5925&  V&  23.59  $\pm$  0.10&   BTA&  This work  \\
5.7072&  r&  23.63  $\pm$  0.08&  NOT&  This work  \\
6.4371&  r&  23.62  $\pm$  0.21&  AZT-20&  This work  \\
6.4555&  R&  23.35  $\pm$  0.22&  AZT-22&  This work  \\
7.4579&  r&  23.69  $\pm$  0.17&  AZT-20&  This work  \\
7.6233&  J&  $>$ 21.80&  SAI-25&  This work  \\
7.7625&  R&  $>$ 24.03&  AZT-22&  This work  \\
8.7252&  u&  25.48$\pm$0.09&  GTC&  This work  \\
8.7252&  g&  24.54$\pm$0.03&  GTC&  This work  \\
8.7252&  r&  24.09$\pm$0.04&  GTC&  This work  \\
8.7252&  i&  24.01$\pm$0.06&  GTC&  This work  \\
8.7252&  z&  23.72$\pm$0.09&  GTC&  This work  \\
8.7595&  H&  21.52$\pm$0.20&  GTC&  This work  \\
9.7213&  r&  23.96  $\pm$  0.12&  NOT&  This work  \\
13.7801&  r&  24.07  $\pm$  0.22&  NOT&  This work  \\
21.5166& 650MHz& $> 75$&  GMRT&  \citetalias{GCN37468}  \\
21.5166& 1260MHz& $> 52$&  GMRT&  \citetalias{GCN37468}  \\
26.8144\tnote{\dag}&  g&  25.13$\pm$0.11&  GTC&  This work  \\
26.8144\tnote{\dag}&  r&  24.42$\pm$0.09&  GTC&  This work  \\
26.8144\tnote{\dag}&  i&  24.53$\pm$0.16&  GTC&  This work  \\
26.8144\tnote{\dag}&  z&  25.13$\pm$0.28&  GTC&  This work  \\
26.8148&  u&  $>$24.03&  GTC&  This work  \\
30.6745\tnote{\dag}&  r&  24.40$\pm$0.19&  NOT&  This work  \\
37.6434\tnote{\dag}&  i&  24.46$\pm$0.20&  NOT&  This work  \\
53.7992&  Ks&   $>$21.10&  GTC&  This work  \\
59.7570\tnote{\dag}&  Ks&  22.12$\pm$0.29&  VLT&  This work  \\
\hline
\end{longtable}
\end{center}
\end{ThreePartTable}

The informations of all telescopes in Table~\ref{tab:optical_result} are as follows: the 0.76~m Katzman Automatic Imaging Telescope (KAIT; located at Lick Observatory, California, United States); the 0.6~m robotic telescope of Burst Observer and Optical Transient Exploring System (BOOTES-7; located at San Pedro de Atacama, Chile); the 1~m telescope at the Las Cumbres Observatory (LCO; located at Siding Springs Observatory, New South Wales, Australia); the 0.7~m telescope of the Thai Robotic Telescope network (TRT; located at Sierra Remote Observatories, California, United States); the 0.40~m SLT (located at Lulin Observatory, Taiwan); the 2.4~m Gao-Mei-Gu  telescope (GMG; located at the Lijiang Observatory, Yunnan, China); the 1.0~m JinShan 100C telescope (ALT/100C; located at Altay Observatory, Xinjiang, China); the 1.5~m AZT-22 telescope of Maidanak Astrophysical Observatory (located at  Qashqadaryo Viloyati, Uzbekistan);
the 2.6~m ZTSh  telescope of the Crimean Astrophysical Observatory (located in Crimea); the 0.7~m AS-32 telescope of the Abastumani Observatory (located at  Abastumani-Kanobili, Georgia); the 2.56~m Nordic Optical Telescope (NOT; located at the Roque de los Muchachos Observatory, La Palma, Spain); the 10.4~m Gran Telescopio Canarias (GTC; Roque de los Muchacos observatory, La Palma, Spain); the 10~m Keck I telescope (located at Maunakea, Hawaii, United States); the 6~m BTA telescope (located at the Special Astrophysical Observatory, Karachay-Cherkessia, Russia); the
1.5~m AZT-20 telescope of Assy-Turgen Observatory (located at  Almaty, Kazakhstan); the 2.5~m SAI-25 alt-azimuth reflector at the Caucasian Mountain Observatory of the Sternberg Astronomical Institute (located at Karachay-Cherkessia, Russia) and the 8.2~m Very Large Telescope (VLT; located at Paranal Observatory, Antofagasta Region, Chile).

\end{document}